\documentstyle{article}
\def\abs#1{\left|{#1}\right|}

\def\mt{{\ifmmode\td M_t\else $\td M_t$\fi}}
\def\as{{\ifmmode\alpha_s\else$\alpha_s$\fi}}
\def\G{\Gamma^{\vphantom{*}}}
\let\lam=\lambda
\let\td=\tilde
\def\cqqh{\hat{c}_{\mathstrut qq}{}}
\def\cqlh{\hat{c}_{\mathstrut ql}{}}
\def\cqlw{\check{c}_{\mathstrut ql}{}}
\def\cudh{\hat{c}_{\mathstrut ud}^*{}}
\def\cueh{\hat{c}_{\mathstrut ue}^*{}}
\def\co#1{{\ifmmode{\cal O}_{#1}\else${\cal O}_{#1}$\fi}}
\def\cs#1{{\ifmmode{\cal S}_{#1}\else${\cal S}_{#1}$\fi}}
\def\at{{\ifmmode{\tilde A}\else$\tilde A$\fi}}
\let\nn=\nonumber

\def\fr#1.#2.{{#1\over #2}}
\let\FR =\fr
\let\CR =\cr
\def\cjkl{C_{ijkl}}
\def\mg{{\ifmmode M_{GUT}\else $M_{GUT}$\fi}}
\begin{document}
\begin{titlepage}
\hfill OHSTPY-HEP-T-96-030 \\

\begin{center} {\large\bf Nucleon Decay in a Realistic SO(10) SUSY GUT}
\vskip .25 in {{Vincent Lucas} and {Stuart Raby}
\\ \vskip .125 in {\footnotesize\sf The Ohio State University, Department of
Physics, 174 W. 18th Ave., Columbus, OH 43210}}
\end{center}

\begin{abstract} In this paper, we calculate neutron and proton decay rates and
branching ratios  in a predictive  SO(10) SUSY GUT which agrees well with low
energy data.  We show that the nucleon lifetimes are consistent with the
experimental bounds. The nucleon decay rates are calculated using all one-loop
chargino and gluino dressed diagrams regardless of their chiral  structure. We
show that the four-fermion operator
$C_{jk} (u_R {d_j}_R)({d_k}_L {\nu_\tau}_L)$, commonly neglected in previous
nucleon decay calculations, not only contributes significantly to nucleon
decay, but, for many values of the initial GUT parameters and for large
$\tan\beta$, actually dominates the decay rate. As a consequence, we find that
$\tau_p/\tau_n$ is often substantially larger than the prediction obtained in
small $\tan\beta$ models.   We also find that gluino-dressed diagrams, often
neglected in nucleon decay calculations, contribute significantly to nucleon
decay. In addition we find that the branching ratios obtained from this
realistic SO(10) SUSY GUT differ significantly from the predictions obtained
from ``generic" SU(5) SUSY GUTS.  Thus nucleon decay branching ratios, when
observed, can be used to test theories of fermion masses.
\end{abstract}
\end{titlepage}

\section{Introduction}

It can easily be seen that Grand Unified Theories [GUTS] contain baryon number
violating operators that produce proton decay \cite{theprotondecays}. However,
determining the decay rate is a much more involved task.  In supersymmetric
[SUSY] GUTS, the nucleon decay amplitude is directly proportional to the
inverse of an effective color triplet Higgs mass \mt\ (resulting from effective
dimension 5 baryon number violating operators) \cite{dim5ops,effcolorhiggs}
multiplied by a product of the Yukawa couplings of this color triplet Higgs to
quarks and leptons. Thus, obtaining a theoretical prediction for nucleon decay
rates depends critically on two factors:
\begin{enumerate}
\item obtaining bounds on the effective color triplet mass, and  

In this regard, it has been noted that \mt\ also affects the prediction for
\as\ through threshold corrections at \mg . Thus, bounds on \mt\ can be
obtained via the experimental constraint on \as\ \cite{hisano,bmp,paper1}, but
only if one has a complete SUSY GUT, valid above \mg .   

\item predictions for the relevant Yukawa couplings. 

The nucleon decay branching ratios are sensitive to these color triplet
Higgs-quark-quark and Higgs-quark-lepton Yukawa couplings.  These couplings are
completely determined in any {\em predictive theory} of fermion masses and
mixing angles, but they are typically {\bf NOT} identical to the Yukawa
couplings responsible for quark and lepton masses. 
\end{enumerate}

In addition, one must take these effective dimension 5 operators and
renormalize them from \mg\ to $M_Z$.  Then, at the weak scale, effective
dimension 6 operators are obtained by closing the squark and/or slepton lines
into  a loop via chargino or gluino exchanges.  Hence the decay rate depends
sensitively on soft SUSY breaking parameters.  Finally the effective dimension
6 operators are renormalized from $M_Z$ to the nucleon mass and then an
effective chiral Lagrangian analysis is used to obtain lifetimes and branching
ratios.

In a recent paper \cite{paper1} [paper I], we proposed several complete SO(10)
SUSY GUTs, valid above \mg.  At \mg, these models had the desired feature that
they reproduced the effective fermion mass operators of model 4 of Anderson, et
al. \cite{adhrs} [paper II].\ \ Moreover, we found that if we required that our
models contain in their superpotentials all terms not forbidden by the
symmetries of the respective models, one and only one additional effective
fermion mass operator was generated for two of the models, while no additional
effective fermion mass generating operators were generated in a third model. We
showed that for reasonable values of the GUT scale parameters, we can obtain
values of
\as\ consistent with experiment, with the
\it effective \rm color triplet Higgs mass that enters into proton decay rates 
$\mt$ as large as $10^{19}$~GeV or even larger.   We argue, however, that  the
scale $\sim 10^{19}$~GeV is, in fact, a natural upper bound for \mt .

Note,  the \it actual \rm color triplets' masses are of order the GUT scale. 
Such a large value for \mt\ in comparison to the GUT scale is obtained in our
model using the Dimopoulos-Wilczek mechanism for doublet-triplet splitting
\cite{DWmech}.  In Appendix 1, we review how the Dimopoulos-Wilczek mechanism
can be used to obtain $\mt \gg M_{GUT}$.  

In this paper, we present the details of our calculations of the nucleon decay
rates for one of our models, referred to as model 4(c), which includes an
additional ``13" mass operator.  In a forthcoming paper, Bla\v zek, Carena,
Wagner, and one of us (S.R.) will show that model 4(c) fits all low energy data
to within 1$\sigma$, while model 4 (without the additional operator) agrees
with all data at the 2$\sigma$ level only \cite{blazek}.   Our main results
for the predictions of the proton and neutron decay rates and branching ratios
in model 4(c) are found in Tables A1A, A1B and A6.

In addition we have studied the sensitivity of our predictions to different
factors.  We have compared the predictions for our model 4(c) to those of
models 4(a) through (f) of Anderson et al.  This tests the sensitivity of the
predicted branching ratios to the quality of the fit for fermion masses and
mixing angles.  We find that branching ratios can differ by factors as large
as 10. 

We have also compared the predictions of our models with those of small
$\tan\beta$ minimal SUSY SU(5) GUTs.  We find that some results in the large
$\tan\beta$ regime are qualitatively different than for small $\tan\beta$.  For
example, certain dimension 6 operators with the chiral structure\footnote{
LLLL, LLRR, and RRRR refer to four-fermion operators pairing  four left-handed
Weyl fermions, pairing two left handed Weyl fermions and the conjugates of two
right-handed Weyl fermions, and pairing the conjugates of four right-handed
Weyl fermions, respectively. Specific examples of each type of four-fermion
operator can be found in Table 3, infra.}
  LLRR tend to dominate over their LLLL counterparts for $\mu(M_Z)/m_{1/2} >
1$.  In this limit, the ratio $\tau_p/\tau_n$, for example, is sensitive to the
quality of the fermion mass fit and is substantially larger than what it is in
small $\tan\beta$ SUSY GUTs.   

Finally we discuss the sensitivity of our predictions to neglecting either the
gluino or chargino exchange diagrams.  We find both contributions to be
significant.

The paper is organized as follows ---  in sections 2, 3 and 4 we discuss the 
calculational ingredients. Namely, in section 2, we discuss GUT models and the
physics from \mg\ to
$M_Z$; in section 3, SUSY loops at $M_Z$ and the resulting dimension 6 baryon
violating operators; and in section 4, the physics from $M_Z$ to the nucleon
mass and a summary of the numerical procedure and our results.  In section 5,
we discuss our results and the aforementioned sensitivities to different
factors.

\section{The low energy effective operators generating nucleon decay, and their
renormalization}

Recall that in paper II, it was shown that four effective fermion mass
operators, denoted \co{33}, \co{23}, \co{12}, and \co{22}, could be used to fit
fermion masses and mixing angles reasonably well. For model 4 of that paper,
the first three of these operators were
\begin{eqnarray}
\co{33} & = & 16_3 \ 10_1 \ 16_3 \nn\\
\co{23} & = & 16_2 \ {A_2\over\at}\  10_1 \ {A_1\over\at}\  16_3 \nn\\
\co{12} & = & 16_1 \ ({\at\over\cs{M}})^3 \ 10_1 \ ({\at\over\cs{M}})^3\  16_2
\nn 
\end{eqnarray} where $16_1$, $16_2$, and $16_3$ are 16 representations
containing the first, second, and third generations of fermions of the Standard
Model, respectively;
$10_1$ is a 10 representation containing the two Higgs doublets of the Minimal
Supersymmetric extension of the Standard Model; $A_1$, $A_2$, and
\at\ are 45s which get vevs in the B-L, hypercharge, and X (SU(5) invariant)
directions, respectively; and \cs{M}\ is an SO(10) singlet getting a vev of
order
$M_{Planck}$. There were six different choices for the operator
\co{22}, labeled a through f.
\begin{eqnarray}
\co{22} = \nn\\ & (a) & 16_2 \ \fr\at.\cs{M}. \  10_1  {A_1 \over \at}  16_2
\nn\\ & (b) &16_2 \ {\cs{G} \over \at}\  10_1 \ \fr A_1.\cs{M}. \ 16_2\nn\\ &
(c) &16_2 \ \fr\at.\cs{M}. \ 10_1 \ \fr A_1.\cs{M}. \ 16_2\nn\\ & (d) &16_2 \
10_1 \ {A_1\over \at}\  16_2\nn\\ & (e) &16_2 \ 10_1 \ \fr\at \ A_1.\cs{M}^2. \
16_2\nn\\ & (f) &16_2 \ 10_1 \ {A_1 \cs{G}\over \at^2}\ 16_2\nn 
\end{eqnarray} where \cs{G} is an SO(10) singlet getting a vev of order
$M_{GUT}$. In addition, from paper I, we discovered that when we built a
complete GUT, valid up to energies around
$M_{Planck}$ and we allowed all terms in the superspace potential consistent
with the symmetries of the theory, an additional fermion mass generating
operator,
\co{13}, was generated for certain versions\footnote{Note, models d, e and f
have the second family $16_2$ coupled directly to $10_1$ and a heavy $16$.  If
this coupling is as large the third generation Yukawa coupling, then we would
obtain excessively large flavor changing neutral current processes, such as
$\mu
\rightarrow e +
\gamma$.} of model 4. 
\begin{eqnarray} \co{13} = \nn\\ & (a) & 16_1\  ({\at \over
\cs{M}})^3\  10_1\  ({\at A_2\over{\cs{M}}^2})\  16_3 \nn\\ & (b) & 0 \nn\\ &
(c) & 16_1
\ ({\at \over \cs{M}})^3 \ 10_1 \ ({A_2\over \cs{M}})\  16_3 \nn \end{eqnarray}

At low energies, these five operators produce effective operators which
generate the fermion masses and which are responsible for baryon-number
violating nucleon decay.
\begin{eqnarray} &\co{33}+\co{23}+\co{22}+\co{12}+\co{13} \nn\\&
\Longrightarrow
\nn\\
 &H_u Q Y_u \overline U  + H_d Q Y_d \overline D + H_d L Y_e
\overline E+ Q {1\over 2}c_{qq} Q T+Q c_{ql} L \overline T+\overline U c_{ud}
\overline D  \overline T+\overline U c_{ue}
\overline E T \nn\\
 & \Longrightarrow \nn\\
 &H_u Q Y_u \overline U + H_d Q Y_d \overline D + H_d L Y_e \overline E+
{1\over{\mt}^{\phantom{(}}}Q {1\over 2}c_{qq} Q \ Q c_{ql}
L+{1\over{\mt}^{\phantom{(}}}\overline U c_{ud}
\overline D 
\ \overline U c_{ue} \overline E \nn
\end{eqnarray} with $T$ and $\overline T$ being the color triplet Higgses from
$10_1$. $Y_u$,
$Y_d$, $Y_e$,
$c_{qq}$, $c_{ql}$, $c_{ud}$, and $c_{ue}$ are flavor matrices which can be
expressed in terms of seven independent real parameters.  At the GUT scale, the
values of these matrices are
$$Y_u=\pmatrix{ 0& C & u_u D e^{i \delta} \CR C & 0 & -\FR1.3. B \CR u'_u D
e^{i
\delta} & -\FR4.3. B & A}$$
$$Y_d=\pmatrix{ 0& -27 C & u_d D e^{i \delta} \CR -27 C & E e^{i \phi} &
\FR1.9. B \CR u'_d D e^{i \delta} & -\FR2.9. B & A}$$
$$Y_e=\pmatrix{ 0& -27 C & u_e D e^{i \delta} \CR -27 C & 3 E e^{i \phi} & B
\CR u'_e D e^{i \delta} & 2 B & A}$$
$$c_{qq}=\pmatrix{ 0 & C & u_{qq} D e^{i \delta} \cr C & y_{qq} E e^{i \phi} &
{1\over3}B \cr u_{qq} D e^{i \delta} & {1\over3}B & A}$$
$$c_{ql}=\pmatrix{ 0 & -27 C & u_{ql} D e^{i \delta} \cr -27 C & y_{ql} E e^{i
\phi} & {1\over3}B \cr u'_{ql} D e^{i \delta} & {1\over3}B & A}$$
$$c_{ud}=\pmatrix{ 0 & -27 C & u_{ud} D e^{i \delta} \cr -27 C & y_{ud} E e^{i
\phi} & -{4\over9}B \cr u'_{ud} D e^{i \delta} & {2\over9}B & A}$$
\begin{equation}
\label{e:matrices} c_{ue}=\pmatrix{ 0 & C & u_{ue} D e^{i \delta} \cr C &
y_{ue} E e^{i \phi} & -4 B \cr u'_{ue} D e^{i \delta} & -2 B & A}
\end{equation} with the values of the Clebsches given in Tables
\ref{t:yclebsch} and
\ref{t:uclebsch}.
\protect
\begin{table}
\vrule height 1pt width 4.75in
\vskip -10pt
\vrule height 1pt width 4.75in
\caption{$y$ Clebsches for each version of model 4}
\label{t:yclebsch}
$$\begin{array}{|c|rrrr|}
\hline
\rm model & y_{qq} & y_{ql} & y_{ud} & y_{ue} \\
\hline a & -3/4 & 3/4 & -5/4 & -3/4 \\ b & -3/2 & 5/2 & 1/2 & -3/2 \\ c & -1/2
& 3/2 & -1/2 & -1/2 \\ d & 3/2 & 3/2 & -1/2 & 3/2 \\ e & 1/2 & 5/2 & 1/2 & 1/2
\\ f & 9/4 & 3/4 & -5/4 & 9/4 \\
\hline
\end{array}$$
\vrule height 1pt width 4.75in
\vskip -10pt
\vrule height 1pt width 4.75in
\end{table}
\begin{table}
\vrule height 1pt width 4.75in
\vskip -10pt
\vrule height 1pt width 4.75in
\caption{$u$ Clebsches for models 4(a) through (c)}
\label{t:uclebsch}
$$\begin{array}{|c|rrrrrrrrrrrrr|}
\hline
\rm model & u_u & u'_u & u_d & u'_d & u_e & u'_e & u_{qq} & u_{ql} & u'_{ql} &
u_{ud} & u'_{ud} & u_{ue} & u'_{ue} \\
\hline a & -4/3 & 1/3 & -2 & -9 & -54 & 3 & 1/3 & 3 & -9 & -2 & 36 & 2 & -4/3
\\ b & 0&0&0&0&0&0&0&0&0&0&0&0&0\\ c & -4/3 & 1/3 & 2/3 & -9 & -54 & -1 & 1/3
& -1 & -9 & 2/3 & 36 & 2 & -4/3 \\
\hline
\end{array}$$
\vrule height 1pt width 4.75in
\vskip -10pt
\vrule height 1pt width 4.75in
\end{table}

These matrices need to be renormalized from the GUT scale to the electroweak
scale. Due to the no-renormalization theorems of supersymmetry, only
wavefunction renormalizations enter into the calculation of the renormalization
group equations of these matrices. The RGEs for the matrices are
\let\dg=\dagger
\def\yu{\ifmmode Y_u^{\phantom{\dg}}\else$Y_u^{\phantom{\dg}}$\fi}
\def\yd{\ifmmode Y_d^{\phantom{\dg}}\else$Y_d^{\phantom{\dg}}$\fi}
\def\ye{\ifmmode Y_e^{\phantom{\dg}}\else$Y_e^{\phantom{\dg}}$\fi}
\def\hyu{\ifmmode Y_u^\dg\else$Y_u^\dg$\fi}
\def\hyd{\ifmmode Y_d^\dg\else$Y_d^\dg$\fi}
\def\hye{\ifmmode Y_e^\dg\else$Y_e^\dg$\fi}
\protect
\begin{eqnarray} {d c_{qq}\over dt} &=& {1\over16\pi^2}\big[ (\yu \hyu+\yd
\hyd) c_{qq} + c_{qq} (\yu \hyu+\yd \hyd)^T \\ &&-({16\over3} g_3^2+3
g_2^2+{1\over15} g_1^2) c_{qq} \big] \nn\\ {d c_{ql}\over dt} &=&
{1\over16\pi^2}\big[ (\yu
\hyu+\yd \hyd) c_{ql} +  c_{ql} (\ye \hye)^T \nn\\ &&-({8\over3} g_3^2+3
g_2^2+{1\over3} g_1^2) c_{ql} \big] \nn\\ {d c_{ud}\over dt} &=&
{1\over16\pi^2}\big[ 2 (\hyu \yu)^T c_{ud} + 2 c_{ud}
\hyd \yd -({16\over3} g_3^2+{2\over3} g_1^2) c_{ud} \big] \nn\\ {d c_{ue}\over
dt} &=& {1\over16\pi^2}\big[ 2 (\hyu \yu)^T c_{ue}+2 c_{ue} \hye
\ye -({8\over3} g_3^2+{26\over15} g_1^2) c_{ue} \big] \nn
\end{eqnarray} with $t=\log (\mu/M_Z)$.

\section{Nucleon decay formulas}

At low energies, the matrices $Y_u$, $Y_d$, and $Y_e$ are diagonalized by
unitary matrices called $S$ and $T$ defined so that
$$S_u Y_u T_u = \pmatrix{
\lam_u\cr &\lam_c\cr &&\lam_t} \equiv \hat Y_u,$$
\let\dag = \dagger and so forth, with all diagonal entries in $\hat Y_u$, $\hat
Y_d$, and $\hat Y_e$ being real and positive. The weak eigenstate basis for the
fermionic fields will be represented by primed fields, and the mass eigenstate
basis will be represented by unprimed letters. Additionally, we choose an
unprimed basis for neutrinos such that 
$$S_e^* \nu'=\nu,$$ so that the lepton-lepton-weak boson vertices are flavor
diagonal in the unprimed basis.\footnote{If left-handed neutrinos have a mass,
the
$\nu_e$, $\nu_\mu$, and
$\nu_\tau$ thus defined are generally not mass eigenstates. Since we sum decay
rates over all neutrino species in our analysis, the convention for defining
the neutrino basis is irrelevant for our results.  In paper I, the 3
left-handed neutrinos remain massless, but this result is easily changed with
the introduction of GUT scale majorana neutrino masses for heavy singlet
neutrinos.}

Squarks and sleptons mass matrices are diagonalized using 6 by 6 matrices $\G$
defined
$$\G_\Omega \left( \begin{array}{c|c} S^*_\Omega \\
\hline & T_\Omega^{T^{\phantom{*}}}
\end{array}\right) \pmatrix{\tilde\Omega' \cr \cr
\smash{\tilde{\overline\Omega}}'^*} =\tilde\Omega, \qquad \Omega=u,d,e$$ where
$\tilde\Omega$ is a six dimensional vector of mass eigenstates and
$\tilde\Omega'$ and $\smash{\tilde{\overline\Omega}}'$ are weak eigenstates.
Cf. notation of \cite{borz}. Additionally, we define $\G_{\Omega,L}$ and
$\G_{\Omega,R}$ to be 6 by 3 matrices so that $\G_{\Omega,L}$ consists of the
first three columns of
$\G_\Omega$ and $\G_{\Omega,R}$ consists of the last three columns. In block
matrix notation,
$$\G_\Omega=\left( \begin{array}{c|c}
\G_{\Omega,L} & \G_{\Omega,R}
\end{array} \right)$$

Since there is no left-handed right-handed neutrino mixing, we will define
$\G_\nu$ so that
$$\G_\nu S_e^* \tilde\nu'=\tilde\nu$$ with $\td\nu$ being a mass eigenstate
basis for the sneutrinos. A right-handed
$\G_\nu$ matrix will not be defined.

Finally, the chargino mass matrices are diagonalized by matrices $U_+$ and
$U_-$ defined 
$$\pmatrix{\td W_+\cr \td H_+}=U_+ \pmatrix{\td\chi_1^+ \cr \td\chi_2^+}$$ and
$$\pmatrix{\td W_-\cr \td H_-}=U_- \pmatrix{\td\chi_1^- \cr \td\chi_2^-}$$
where
$\td\chi^\pm_{1,2}$ are mass eigenstates.

In this notation, various Feynman vertices relevant to nucleon decay are given
in Appendix 2.

\subsection{Gluino diagrams}

Figure \ref{f:1} shows the two diagrams which contribute to the four-fermion
operator $\cjkl^{(ud)(d\nu)[G]}\* (u_i^\alpha d_j^\beta) (d_k^\gamma \nu_l)
\epsilon_{\alpha\beta\gamma}$. Calculating these two diagrams, we find
\protect
\begin{eqnarray} & C_{ijkl}^{(ud)(d\nu)[G]}={4\over3}{1\over16\pi^2 \mt}g_3^2
\bigl\{
\G_{U,L \; \lambda i}
\G{}^*_{U,L \; \lambda i'} \G_{D,L \; \rho j} \G{}^*_{D,L \; \rho j'}
\cqqh^{i'[j'} \cqlh^{k] l} m_{\td{g}} I(\td g,\td u_\lam,\td d_\rho)-\bigr.
\nn\\ &\bigl. \G_{D,L \; 
\rho j}
\G{}^*_{D,L \; \rho j'} \G_{D,L \; \sigma k} \G{}^*_{D,L \; \sigma k'}
\cqqh^{i[j'} \cqlh^{k'] l} m_{\td{g}} I(\td g,\td d_\rho, \td d_\sigma)\bigr\}
\nn
\end{eqnarray} where $\alpha, \beta,$ and $\gamma$ are color indices, 
$i$, $i'$, $j$, $j'$,
$k$, and $l$ are fermion flavor indices, $\rho$ and $\lam$ are squark flavor
indices, $\cqqh=S_u c_{qq}^R S_d^T$,
$\cqlh=S_d c_{ql}^R S_e^T$ with $c_{qq}^R$ and $c_{ql}^R$ being $c_{qq}$ and
$c_{ql}$, respectively, renormalized to $M_Z$, and
$$I(a,b,c)=\def\mac#1#2#3{{#1^2 \log #1^2 \over(#1^2-#2^2)(#1^2-#3^2)}}
\mac{m_a}{m_b}{m_c}+\mac{m_b}{m_c}{m_a}+\mac{m_c}{m_a}{m_b}$$ where $m_a$
equals the mass of particle $a$, etc.\footnote{We use a notation for
antisymmetrization on tensor indices in which
$A_{i[jkl]mn}$ is equal to $A_{ijklmn}-A_{ilkjmn}$, and so forth.}
\def\tri #1,#2,#3,#4,#5,#6,#7.{
\begin{picture}(150,150)(0,-75)
\thicklines
\put(0,50){\line(1,0){50}}
\put(25,50){\vector(1,0){0}}
\put(0,-50){\line(1,0){50}}
\put(25,-50){\vector(1,0){0}}
\put(150,50){\line(-1,-1){50}}
\put(150,-50){\line(-1,1){50}}
\multiput(50,50)(20,-20){3}{\line(1,-1){10}}
\multiput(50,-50)(20,20){3}{\line(1,1){10}}
\put(50,50){\line(0,-1){100}}
\put(50,25){\vector(0,1){0}}
\put(50,-25){\vector(0,-1){0}}
\put(78,22){\vector(1,-1){0}}
\put(78,-22){\vector(1,1){0}}
\put(125,25){\vector(-1,-1){0}}
\put(125,-25){\vector(-1,1){0}}
\put(18,60){$#1$}
\put(18,-70){$#2$}
\put(82,32){$#5$}
\put(82,-40){$#6$}
\put(135,-25){$#4$}
\put(135,17){$#3$}
\put(32,-6){$#7$}
\end{picture}}
\begin{figure}
\vrule height 1pt width 4.75in
\vskip -10pt
\vrule height 1pt width 4.75in
\begin{center}
\caption{}
\label{f:1}
\vbox{
\tri  u_i^\alpha,d_j^\beta,d_k^\gamma,\nu_l,\td u_\lambda^{\alpha'},
\td d_\rho^{\beta'},\td g. }
\vskip .5 in
\tri d_j^\beta,d_k^\gamma,u_i^\alpha,\nu_l,\td d_\rho^{\beta'},
\td d_\sigma^{\gamma'},\td g.
\end{center}
\vrule height 1pt width 4.75in
\vskip -10pt
\vrule height 1pt width 4.75in
\end{figure}

By similar calculations, we can construct the remaining four-fermion operators
listed in Table \ref{t:operators}. Note, we also use Table \ref{t:operators} to
define our notation for these four-fermion operators.  The $\cjkl$s for  these
operators are
$$\cjkl^{(ud)(ue)[G]}=-{4\over3}{1\over16\pi^2\mt}g_3^2 \bigl\{ \G_{U,L \;
\lambda i}
\G{}^*_{U,L \; \lambda i'} \G_{D,L \; \rho j} \G{}^*_{D,L \; \rho j'}
\cqqh^{[i'j'} \cqlw^{k] l} m_{\td{g}} I(\td g,\td u_\lam, \td d_\rho)-\bigr.$$
$$\bigl. \G_{U,L \; \lam i}
\G{}^*_{U,L \; \lam i'} \G_{U,L \; \sigma k} \G{}^*_{U,L \; \sigma k'}
\cqqh^{[i'j} \cqlw^{k'] l} m_{\td{g}} I(\td g,\td u_\lam, \td u_\sigma)\bigr\}
$$
$$\cjkl^{(\overline{dd})(u\nu)[G]}={4\over3}{1\over16\pi^2\mt}g_3^2 \G_{D,R \; 
\rho j}
\G{}^*_{D,L \; \rho j'} \G_{D,R \; \sigma k} \G{}^*_{D,L \; \sigma k'}
\cqqh^{i[j'} \cqlh^{k'] l} m_{\td{g}} I(\td g,\td d_\rho, \td d_\sigma)
$$
$$\cjkl^{(\overline{ud})(d\nu)[G]}= {4\over3}{1\over16\pi^2\mt}g_3^2 \G_{U,R \;
\lambda i}
\G{}^*_{U,L \; \lambda i'} \G_{D,R \; \rho j} \G{}^*_{D,L \; \rho j'}
\cqqh^{i'[j'} \cqlh^{k] l} m_{\td{g}} I(\td g,\td u_\lam, \td d_\rho)
$$
$$\cjkl^{(\overline{ud})(ue)[G]}= -{4\over3}{1\over16\pi^2\mt}g_3^2
\G_{U,R \; \lambda i}
\G{}^*_{U,L \; \lambda i'} \G_{D,R \; \rho j} \G{}^*_{D,L \; \rho j'}
\cqqh^{[i'j'} \cqlw^{k] l} m_{\td{g}} I(\td g,\td u_\lam, \td d_\rho)
$$
$$\cjkl^{(ud)(\overline{ue})[G]}= -{4\over3}{1\over16\pi^2\mt}g_3^2
\G_{U,L \; \lambda i}
\G{}^*_{U,R \; \lambda i'} \G_{D,L \; \rho j} \G{}^*_{D,R \; \rho j'}
\cudh^{[i'j'} \cueh^{k] l} m_{\td{g}} I(\td g,\td u_\lam, \td d_\rho)
$$ 
$$\cjkl^{(\overline{\mathstrut ud})(\overline{\mathstrut ue})[G]}=
-{4\over3}{1\over16\pi^2\mt}g_3^2 \bigl\{\G_{U,R \; \lambda i}
\G{}^*_{U,R \; \lambda i'} \G_{D,R \; \rho j} \G{}^*_{D,R \; \rho j'}
\cudh^{[i'j'} \cueh^{k] l} m_{\td{g}} I(\td g,\td u_\lam, \td d_\rho)-\bigr.$$
$$\bigl. \G_{U,R \; \lam i}
\G{}^*_{U,R \; \lam i'} \G_{U,R \; \sigma k} \G{}^*_{U,R \; \sigma k'}
\cudh^{[i'j} \cueh^{k'] l} m_{\td{g}} I(\td g,\td u_\lam, \td u_\sigma)\bigr\}
$$ where $\cqlw=S_u c_{ql}^R S_e^T$, $\hat c_{ud}=T_u^T c_{ud}^R T_d$, and 
$\hat c_{ue}=T_u^T c_{ue}^R T_e$ with $c_{ud}^R$ and $c_{ue}^R$ being $c_{ud}$
and
$c_{ue}$, respectively, renormalized to $M_Z$.
\let\ep=\epsilon
\def\abg{{\alpha \beta \gamma}}
\protect
\begin{table}
\vrule height 1pt width 4.75in
\vskip -10pt
\vrule height 1pt width 4.75in
\begin{center}
\caption{Table of all gluino-dressed four fermion operators relevant to nucleon
decay}
\label{t:operators}
$$\hskip -.75 in
\begin{array}{|c|c|c|}
\hline
\multicolumn{3}{|c|}{\rm operator\ type} \\
\hline
\rm LLLL & \rm LLRR & \rm RRRR \\
\hline
\cjkl^{(ud)(d\nu)[G]}\* (u_i^\alpha d_j^\beta) (d_k^\gamma \nu_l)
\epsilon_{\alpha\beta\gamma} &\cjkl^{(\overline{dd})(u\nu)[G]}\* (\overline
d_j^{* \beta} \overline d_k^{* \gamma}) (u_i^\alpha \nu_l) \ep_\abg &\\
&\cjkl^{(\overline{ud})(d\nu)[G]}\* (\overline{\mathstrut u}_i^{* \alpha} 
\overline{\mathstrut d}_j^{* \beta}) (d_k^\gamma \nu_l) \ep_\abg&\\
\hline
\cjkl^{(ud)(ue)[G]}\* (u_i^\alpha d_j^\beta) (u_k^\gamma e_l)
\epsilon_{\alpha\beta\gamma} &\cjkl^{(\overline{ud})(ue)[G]}\*
(\overline{\mathstrut u}_i^{* \alpha} 
\overline{\mathstrut d}_j^{* \beta}) (u_k^\gamma e_l) \ep_\abg
&\cjkl^{(\overline{\mathstrut ud})(\overline{\mathstrut ue})[G]}\*
(\overline{\mathstrut u}_i^{* \alpha}
\overline{\mathstrut d}_j^{* \beta}) (\overline{\mathstrut u}_k^{* \gamma}
\overline{\mathstrut e}_l^*)
\epsilon_{\alpha
\beta \gamma}\\ &\cjkl^{(ud)(\overline{ue})[G]}\* (u_i^\alpha d_j^\beta) 
(\overline{\mathstrut u}_k^{* \gamma} 
\overline{\mathstrut e}_l^*) \ep_\abg&\\
\hline
\end{array}$$
\end{center}
\vrule height 1pt width 4.75in
\vskip -10pt
\vrule height 1pt width 4.75in
\end{table}

As observed in ref. \cite{gluinosarezero}, the contribution of gluino dressed
operators is zero in the limit that all squarks are degenerate at the
electroweak scale.  In this analysis, however, squarks and sleptons are assumed
degenerate at \mg\ and as a consequence of renormalization group running they
are explicitly non-degenerate at the weak scale.  Thus we retain the gluino
contribution in our analysis.

In principle, we could also have four-fermion operators like 
$\cjkl^{(\overline{uu})(de)[G]}\* (\overline u_i^{* \alpha} \overline u_k^{*
\gamma}) (d_j^\beta e_l) \ep_\abg$ pairing two up type quarks in a Weyl index
contracted pair.  However, since we are only interested in the first generation
of up quarks, the portions of these operators that would be relevant to nucleon
decay are identically zero.

\subsection{Chargino diagrams}

\def\erase#1{{\smash{\phantom{#1}}}}
\def\gu{[(g_2 \G_{U,L} U_{+ \; 1n}-\G_{U,R} \hat Y_u^\erase{T} U_{+\; 2n})
V_{KM}]}
\def\gd{[(g_2 \G_{D,L} U_{- \; 1n}-\G_{D,R} \hat Y_d^\erase{T} U_{-\; 2n})
V_{KM}^\dg]}
\def\ge{[(g_2 \G_{E,L} U_{- \; 1n}-\G_{E,R} \hat Y_e^\erase{T} U_{-\; 2n})]}
\def\gub{U^*_{- \; 2n} (\G_{U,L} V_{KM} \hat Y_d^\erase{*})}
\def\gdb{U^*_{+ \; 2n} (\G_{D,L} V_{KM}^\dg \hat Y_u^\erase{*})}
\def\gnub{U^*_{- \; 2n} (\G_{\nu} \hat Y_e^\erase{*})} Using the vertices in
Appendix 2, the chargino diagrams can be readily computed. According to
analyses by many other authors on proton decay in SUSY GUTs, the dominant
operators would be expected to be 
$\cjkl^{(ud)(d\nu)[W]}\* (u_i^\alpha d_j^\beta) (d_k^\gamma \nu_l)
\epsilon_{\alpha\beta\gamma}$ and $\cjkl^{(ud)(ue)[W]}\* (u_i^\alpha d_j^\beta)
(u_k^\gamma e_l)
\epsilon_{\alpha\beta\gamma}$. The $\cjkl$s for these operators are equal to 
$$\cjkl^{(ud)(d\nu)[W]}= {1\over16\pi^2\mt}\Big\{ \gd_{\rho i}$$
$$\times \gu_{\lam j}$$
$$\times \G{}^*_{U,L \; \lam i'} \G{}^*_{D,L \; \rho j'}
\cqqh^{i'[j'} \cqlh^{k]l} m_{\td\chi_n} I(\td\chi_n,\td u_\lam,\td d_\rho)$$
$$+
\gu_{\lam k}$$
$$\times \ge_{\rho l}$$
$$\times \G{}^*_{U,L \; \lam k'} \G{}^*_{E,L \; \rho l'}
\cqqh^{[ij} \cqlw^{k']l'} m_{\td\chi_n} I(\td\chi_n,\td u_\lam,\td e_\rho)
\Big\}
$$

\def\xn{{\td\chi_n}}
\def\mr{$$ $$\times}
\def\pr{$$ $$+}
$$\cjkl^{(ud)(ue)[W]}= -{1\over16\pi^2\mt}\Big\{ \gd_{\rho i} \mr
\gu_{\lam j}\mr
\G{}^*_{U,L \; \lam i'} \G{}^*_{D,L \; \rho j'}
\cqqh^{[i' j'} \cqlw^{k]l} m_{\xn} I(\xn,\td u_\lam,\td d_\rho)\pr
\gd_{\rho k}\mr g_2 \G_{\nu \; \lam l} U_{+ \;1n}
\G{}^*_{D,L \; \rho k'} \G{}^*_{\nu \; \lam l'} \cqqh^{i[j}
\cqlh^{k']l'} m_\xn I(\xn,\td d_\rho,\td \nu_\lam) \Big\}$$

The $\cjkl$s for the LLRR and RRRR operators are
$$\cjkl^{(\overline{ud})(d\nu)[W]}= {1\over16\pi^2\mt} \Big\{
\gdb_{\rho i} \gub_{\lam j}\mr
\G{}^*_{U,L \; \lam i'} \G{}^*_{D,L \; \rho j'}
\cqqh^{i'[j'} \cqlh^{k]l} m_\xn I(\xn,\td u_\lam,\td d_\rho)\pr
\gu_{\lam k}\mr
\ge_{\rho l}\mr
\G{}^*_{U,R \; \lam k'} \G{}^*_{E,R \; \rho l'}
\cudh^{[ij} \cueh^{k']l'} m_\xn I(\xn,\td u_\lam,\td e_\rho) \Big\}$$

$$\cjkl^{(ud) (\overline{ue})[W]}= -{1\over16\pi^2\mt}\Big\{ \gd_{\rho i}\mr
\gu_{\lam j}\mr
\G{}^*_{U,R \; \lam i'} \G{}^*_{D,R \; \rho j'}
\cudh^{[i' j'} \cueh^{k]l} m_\xn I(\xn,\td u_\lam, \td d_\rho)\pr
\gdb_{\rho k} \gnub_{\lam l}\mr
\G{}^*_{D,L \; \rho k'} \G{}^*_{\nu \; \lam l'}
\cqqh^{i[j} \cqlh^{k']l'} m_\xn I(\xn,\td d_\rho, \td\nu_\lam)\Big\}$$

$$\cjkl^{(\overline{ud})(ue)[W]}= -{1\over16\pi^2\mt}\Big\{
\gdb_{\rho i} \gub_{\lam j}\mr
\G{}^*_{U,L \; \lam i'} \G{}^*_{D,L \; \rho j'}
\cqqh^{[i' j'} \cqlw^{k]l} m_\xn I(\xn,\td u_\lam,\td d_\rho)\Big\}$$

$$\cjkl^{(\overline{\mathstrut ud})(\overline{\mathstrut ue})[W]}=
-{1\over16\pi^2\mt} \Big\{ \gdb_{\rho i} \gub_{\lam j}\mr
\G{}^*_{U,R \; \lam i'} \G{}^*_{D,R \; \rho j'}
\cudh^{[i' j'} \cueh^{k]l} m_\xn I(\xn,\td u_\lam, \td d_\rho)\Big\}$$

\section{Numerical procedure and results}

In calculating the following nucleon decay rates, we make the standard
universality assumptions about the soft SUSY parameters at \mg, except that we
allow non-universal values for $M_{H_u}$ and $M_{H_d}$.\footnote{Note that if
the messenger scale of SUSY breaking is $M_{Planck}$ then our analysis is not
completely self-consistent.  In any complete SUSY GUT defined up to an
effective cut-off scale $M > M_G$,  the interactions above $M_G$ will
renormalize the soft breaking parameters.   This will, in general, split the
degeneracy of squark and slepton masses at 
$M_G$ even if they are degenerate at $M$.  On the other hand, bounds on flavor
changing neutral current processes, severely constrain the magnitude of 
possible splitting. Thus these corrections must be small. In addition, in 
theories where SUSY breaking is mediated by gauge exchanges  with a messenger 
scale below (but near) $M_G$,  the present analysis is expected to apply 
unchanged.  Since in this case squarks and sleptons will be nearly degenerate 
at the messenger scale.  The Higgs masses, on the other hand, are probably
dominated by new interactions which also generate a $\mu$ term.  It is thus 
plausible to expect the Higgs masses to be split and independent of squark and 
slepton masses.  The parameter $A_0$ could also be universal at the messenger 
scale.} We define \mg\ such that $\alpha_1=\alpha_2\equiv\tilde\alpha_{GUT}$ 
at
\mg\ and define $\ep_3$, representing the contribution of GUT scale threshold
corrections to the gauge couplings, to be
$(\alpha_3(\mg)-\tilde\alpha_{GUT})/\tilde\alpha_{GUT}$. The dimensionless
(dimensionful) parameters are renormalized at two (one) loops to $M_Z$ using
the renormalization group equations of Martin and Vaughn \cite{RGE}, except
that
$c_{qq}$, $c_{ql}$, $c_{ud}$, and $c_{ue}$ are renormalized at one loop using
the equations of this paper.

Renormalization of the $\cjkl$s from $M_Z$ to 1 GeV is taken into account by
multiplying them by the $A_L$ calculated in \cite{al}, and then chiral
Lagrangian techniques \cite{chadha} are used to obtain nucleon decay
amplitudes. Formulas for nucleon decay rates in terms of the chiral Lagrangian
parameters are contained in Appendix 3.

The decay rates depend heavily on the chiral Lagrangian factors $\alpha$ and
$\beta$ where $$\beta U({\bf k}) =\ep_\abg \smash{<}0 | (u^\alpha d^\beta)
u^\gamma |{\rm proton}({\bf k})\smash{>} ,$$  $$\alpha U({\bf k}) =\ep_\abg
\smash{<}0 |(\overline{\vphantom{d}u}^{*\,\alpha}
\overline d^{*\,\beta}) u^\gamma |{\rm proton}({\bf k})\smash{>}$$ and $U({\bf
k})$ is the left handed component of the proton's wavefunction. See, e.g., ref.
\cite{b:eq:a}. It is known that $|\beta|=|\alpha|$ \cite{b:eq:a,gavela} and
that
$|\beta|$ ranges from .003 to .03 GeV${}^3$ \cite{calc:beta}. Lattice
calculations have not reduced the uncertainty in $|\beta|$; lattice
calculations have reported $|\beta|$ as low as .006 GeV${}^3$ \cite{gavela}
and as high as .03 GeV${}^3$ \cite{hara}. Additionally, the phase between
$\alpha$ and $\beta$ is not widely reported, although a relatively recent
lattice calculation suggests that $\beta=-\alpha$ \cite{gavela}. Therefore, we
have left the phase between $\alpha$ and $\beta$ a free variable, and report
three values for many of the quantities predicted in our tables. Namely, the
max (min) referred to in the tables is the value for the quantity predicted
when the phase between
$\alpha$ and $\beta$ is such that the quantity is maximized (minimized). Hence,
each entry in the max and min columns uses a different value of
$\arg(\beta/\alpha)$.

In the following tables, we have calculated decay rates using $\mt=10^{19}$~GeV
and $|\beta|=.003$~GeV${}^3$. In paper I, we showed that without any fine
tuning this value of \mt\ can be made consistent with the measured value for
\as.  In addition, we argued that it seems unnatural to have \mt\ much bigger
than
$10^{19}$~GeV in the sense that in order to have \mt\ much bigger than
$10^{19}$~GeV, there would need to be a supermassive electroweak doublet in the
GUT desert with mass many orders of magnitude lower than the GUT scale itself
and at least an order of magnitude lighter than any other particle getting mass
around the GUT scale. Therefore, the values presented in the tables are roughly
upper bounds on the nucleons' lifetimes, based on naturalness\footnote{A note
of caution -- it was also shown in ref. \cite{gavela} that  chiral Lagrangian
techniques overestimate the amplitude $\ep_\abg \smash{<}\pi^0 | (u^\alpha
d^\beta) u^\gamma |p \smash{>}$ by at least a factor of 2.4 (for $\beta =
0.006$).  Thus the proton decay rate is overestimated by almost a factor of 6
or more in this case. As a result, for any given $|\beta|$, the actual nucleon
lifetimes for that $\beta$, could be a factor of 6 or more larger than the
results reported here, as extrapolated from our tables for that value of
$|\beta|$.  These remarks are indicative of the theoretical uncertainty in the
calculation due to strong interaction effects.}. Since all decay rates scale as
$({|\beta|\over .003{\rm\ GeV^3}} {10^{19}{\rm\ GeV}\over\mt})^2$, the
lifetimes for different values of \mt\ and $|\beta|$ can easily be extracted
from the tables.

In Tables A1 through A6 [the A tables], we calculate nucleon decay using model
4(c) (including the \co{13} operator of paper I), since it appears to be the
model which best fits the low energy data \cite{blazek}. Initial values for the
dimensionless Yukawa parameters; soft SUSY parameters; and \mg,
$\tilde\alpha_{GUT}$, and $\tan \beta$ are taken from the global $\chi^2$
analysis of Bla\v zek et al. \cite{blazek}.  This global $\chi^2$ analysis
shows that these values of the GUT parameters are consistent with electroweak
symmetry breaking and the experimental bounds on the sparticle masses; and
that, for the particular values of the soft SUSY breaking parameters used,
these parameters give the best global fit to 20 low energy observables,
including experimental measurements for the gauge couplings; fermion masses
and mixing angles; and $b
\to s\gamma$, with $\chi^2$ per degree of freedom $< 1\fr 1.3.$.  Tables A1A
and A1B contain the lifetimes for
$p\to K^+\overline\nu$ and $n\to K^0
\overline\nu$ and compares these rates with the rates of decay into the other
significant decay modes involving spin zero mesons, for various values of the
GUT scale parameters. These are the main results of this paper.  Note however
that only the three or four most significant decay modes are included in these
tables.

In the subsequent tables we evaluate the relative contributions to these rates
from different sources -- (LLRR vs. LLLL operators)  or  (gluinos vs.
charginos).  Tables A2A, A2B, A3A, and A3B compare the contributions of
chargino and gluino diagrams to the total decay rate. Table A4 compares the
contribution of LLLL versus LLRR operators for decays into anti-neutrinos, for
each of the three anti-neutrino species. Tables A5A and A5B compare the
relative importance of each generation of anti-neutrino to the total proton
decay rate, for decay modes involving anti-neutrinos. Finally, Table A6
contains the values of the GUT scale parameters used in Tables A1 through
A5.   
 
Tables B1 through B6 [the B tables] contain information similar to that in the
A tables except that they compare models 4(a) through (f), without the
\co{13}\ operator.   We used values for the initial (GUT scale) parameters,
taken from unreported data from the collaboration of ref. \cite{blazek}, which
are consistent with electroweak symmetry breaking and the experimental bounds
on sparticle masses, and which give predictions agreeing to within
$2.1\sigma$ with experimental measurements for gauge couplings, fermion masses
and mixing angles, and $b \to s\gamma$.  This comparison gives us information
on the model dependence of nucleon decay branching ratios.  Note that models
4(a) through (f) (without the \co{13}\ operator) give identical results for
fermion masses and mixing angles.  This is because the contribution of the
different
\co{22} operators to the 22 entry of the Yukawa matrices all give the same
Clebsch relation  0:1:3 for u:d:e matrices \cite{adhrs}.  They however have
different Clebsch relations for the 22 entry of the matrices $c_{qq}$,
$c_{ql}$, $c_{ud}$, and $c_{ue}$ relevant for nucleon decay.  

In addition, the comparison of model 4(c), with the \co{13} operator, to models
4(a) through (f), without the \co{13} operator, gives us information on the
sensitivity of predictions for nucleon decay with respect to the quality of the
fit for fermion masses and mixing angles.   In order to compare  the runs in
the A tables directly with the runs in the B tables, we chose the GUT scale
parameters such that for each run in the B tables, the values for
$\tilde\alpha_{GUT}^{-1}$, $M_{GUT}$,
$\epsilon_3$; the soft SUSY breaking parameters $\tan\beta$, $\mu$, $m_{1/2}$,
$m_0$, $m_{H_u}$, $m_{H_d}$, and $A_0$; and the Yukawa parameter $A$ are nearly
the same as they are for the run of the same name in the A tables.

\def\G{\Gamma}
\def\rr#1.{\G(p \to #1)}
\def\frac#1#2{{#1 \over #2}}
\def\rx#1.{\G_{p \to #1}}
\def\rn#1.{\G(n \to #1)}
\def\rnx#1.{\G_{n \to #1}}
\def\rrx#1.{\G_{p \ \ \to\atop #1}}

\footnotesize
\vbox{
\vskip 11 pt {\hskip -.75 in
\vrule height 1pt width 5.75in
\vskip -7pt
\hskip -.75 in
\vrule height 1pt width 5.75in}
$$\hskip -1.0625 in
\begin{array}{|c|ccc|ccc|ccc|ccc|ccc|}
\hline
\rm{run\atop no.} & \multicolumn{3}{|c|}{\tau (p \to K^+ \overline\nu)/(10^{32}
{\rm yrs})} & \multicolumn{3}{|c|}{\frac{\G(p \to \pi^+ \overline\nu)}{\G(p \to
K^+ \overline\nu)}} &
\multicolumn{3}{|c|}{\frac{\rr K^0 \mu^+.}{\rr K^+ \overline\nu.}\times 10^2} &
\multicolumn{3}{|c|}{\frac{\rr \pi^0 \mu^+.}{\rr K^+ \overline\nu.}\times
10^2}&
\multicolumn{3}{|c|}{\frac{\rr \eta \mu^+.}{\rr K^+ \overline\nu.}\times
10^2}\\
\cline{2-16} & \rm max &\beta=-\alpha & \rm min & \rm max &\beta=-\alpha & \rm
min& \rm max &\beta=-\alpha & \rm min& \rm max &\beta=-\alpha & \rm min& \rm
max &\beta=-\alpha & \rm min\\ 
\hline
 \rm I & 26. & 14. & 14. & 0.46 & 0.39 & 0.39 & 0.52 & 0.29 & 0.29 & 0.28 &
0.16 & 0.16 & 0.099 & 0.055 & 0.055\cr
 \rm II & 60. & 37. & 37. & 0.37 & 0.33 & 0.32 & 0.53 & 0.33 & 0.32 & 0.29 &
0.18 & 0.18 & 0.10 & 0.062 & 0.061\cr
 \rm III(1) & 220. & 130. & 98. & 1.1 & 0.74 & 0.69 & 0.31 & 0.19 & 0.14 &
0.12 & 0.073 & 0.055 & 0.028 & 0.017 & 0.013\cr
 \rm III(2) & 150. & 97. & 74. & 1.5 & 1.1 & 0.94 & 0.37 & 0.24 & 0.18 & 0.11 &
0.071 & 0.054 & 0.017 & 0.011 & 0.0084\cr
 \rm III(3) & 110. & 76. & 58. & 1.5 & 1.1 & 1.0 & 0.34 & 0.23 & 0.18 & 0.092 &
0.063 & 0.049 & 0.011 & 0.0078 & 0.0060\cr
\hline
\end{array}$$ {\small Table A1A: Partial mean lifetime for proton decaying into
kaon plus anti-neutrino and ratios of the rates of proton decay into various
decay products versus rate of decay into kaon plus anti-neutrino for various
values of the GUT scale parameters, when the
\co{13} operator is included.}}

\normalsize
\vbox{$$\hskip -.8 in
\begin{array}{|c|ccc|ccc|ccc|ccc|}
\hline
\rm{run\atop no.}  & \multicolumn{3}{|c|}{\tau (n \to K^0
\overline\nu)/(10^{32} {\rm yrs})}  & \multicolumn{3}{|c|}{\frac{\G(n \to
\pi^0 \overline\nu)}{\G(n \to K^0 \overline\nu)}\times 10^2} &
\multicolumn{3}{|c|}{\frac{\rn \eta\overline\nu.}{\rn K^0 \overline\nu.}\times
10^2} &
\multicolumn{3}{|c|}{\frac{\rn \pi^- \mu^+.}{\rn K^0 \overline\nu.}\times 10^2}
\\
\cline{2-13}  & \rm max &\beta=-\alpha & \rm min & \rm max &\beta=-\alpha & \rm
min& \rm max &\beta=-\alpha & \rm min& \rm max &\beta=-\alpha & \rm min \\ 
\hline
 \rm I & 3.8 & 2.3 & 2.3 & 3.3 & 3.2 & 3.2 & 0.37 & 0.17 & 0.17 & 0.083 & 0.051
& 0.050\cr
 \rm II & 12. & 7.3 & 6.9 & 3.3 & 3.2 & 3.2 & 0.49 & 0.25 & 0.23 & 0.11 & 0.070
& 0.067\cr
 \rm III(1) & 14. & 12. & 9.8 & 3.5 & 3.5 & 3.4 & 0.13 & 0.088 & 0.035 & 0.016
& 0.014 & 0.011\cr
 \rm III(2) & 6.4 & 5.8 & 5.0 & 3.2 & 3.1 & 3.1 & 0.079 & 0.055 & 0.022 &
0.0094 & 0.0085 & 0.0073\cr
 \rm III(3) & 4.5 & 4.2 & 3.7 & 3.2 & 3.1 & 3.1 & 0.063 & 0.045 & 0.019 &
0.0076 & 0.0070 & 0.0061\cr
\hline
\end{array}$$ {\small Table A1B: Partial mean lifetime for neutron decaying
into kaon plus anti-neutrino and ratios of the rates of proton decay into
various decay products versus rate of decay into kaon plus anti-neutrino for
various values of the GUT scale parameters, when the
\co{13} operator is included.}}

\normalsize
\vbox{$$\begin{array}{|c|ccc|ccc|ccc|}
\hline
\rm{run\atop no.} & \multicolumn{3}{|c|}{\frac{\rx K^+ \overline\nu.^{\rm
chargino}}{\rx K^+ \overline\nu.^{\rm total}}} &
\multicolumn{3}{|c|}{\frac{\rx \pi^+ \overline\nu.^{\rm chargino}}{\rx \pi^+
\overline\nu.^{\rm total}}} &
\multicolumn{3}{|c|}{\frac{\rx K^0 \mu^+.^{\rm chargino}}{\rx K^0 \mu^+.^{\rm
total}}} \\
\cline{2-10} & \max & \beta=-\alpha & \min& \max & \beta=-\alpha & \min& \max &
\beta=-\alpha & \min \\
\hline
 \rm I & 1.1 & 1.1 & 0.90 & 1.1 & 1.1 & 0.87 & 0.99 & 0.99 & 0.99\cr
 \rm II & 1.3 & 1.2 & 0.82 & 1.3 & 1.2 & 0.73 & 1.0 & 1.0 & 1.0\cr
 \rm III(1) & 1.9 & 1.4 & 0.45 & 1.4 & 1.3 & 0.71 & 1.2 & 1.2 & 1.2\cr
 \rm III(2) & 1.7 & 1.3 & 0.52 & 1.3 & 1.2 & 0.79 & 1.1 & 1.1 & 1.1\cr
 \rm III(3) & 1.6 & 1.3 & 0.57 & 1.2 & 1.2 & 0.83 & 1.1 & 1.1 & 1.1\cr
\hline
\end{array}$$ {\small Table A2A: Ratios of the rate of proton decay that would
occur if chargino diagrams contributed only versus total proton decay rate for
the three most dominant decay modes, for various values of the GUT scale
parameters, when the \co{13} operator is included.}}

\vbox{$$\hskip -.625 in
\begin{array}{|c|ccc|ccc|ccc|ccc|}
\hline
\rm{run\atop no.} & 
\multicolumn{3}{|c|}{\frac{\rnx K^0 \overline\nu.^{\rm chargino}}{\rnx K^0
\overline\nu.^{\rm total}}} &
\multicolumn{3}{|c|}{\frac{\rnx \pi^0 \overline\nu.^{\rm chargino}}{\rnx \pi^0
\overline\nu.^{\rm total}}} &
\multicolumn{3}{|c|}{\frac{\rnx \eta \overline\nu.^{\rm chargino}}{\rnx \eta
\overline\nu.^{\rm total}}} &
\multicolumn{3}{|c|}{\frac{\rnx \pi^- \mu^+.^{\rm chargino}}{\rnx \pi^-
\mu^+.^{\rm total}}} \\
\cline{2-13} & \max & \beta=-\alpha & \min& \max & \beta=-\alpha & \min& \max &
\beta=-\alpha & \min & \max & \beta=-\alpha & \min\\
\hline
 \rm I & 1.1 & 1.1 & 0.87 & 1.1 & 1.1 & 0.87 & 1.1 & 0.99 & 0.97 & 0.81 & 0.81
& 0.81\cr
 \rm II & 1.3 & 1.2 & 0.73 & 1.3 & 1.2 & 0.73 & 1.2 & 0.99 & 0.96 & 0.64 & 0.64
& 0.64\cr
 \rm III(1) & 1.4 & 1.3 & 0.71 & 1.4 & 1.3 & 0.71 & 2.2 & 0.44 & 0.40 & 0.43 &
0.43 & 0.43\cr
 \rm III(2) & 1.2 & 1.2 & 0.80 & 1.3 & 1.2 & 0.79 & 2.2 & 0.42 & 0.39 & 0.55 &
0.55 & 0.55\cr
 \rm III(3) & 1.2 & 1.1 & 0.84 & 1.2 & 1.2 & 0.83 & 2.2 & 0.43 & 0.41 & 0.63 &
0.63 & 0.63\cr
\hline
\end{array}$$ {\small Table A2B: Ratios of the rate of neutron decay that would
occur if chargino diagrams contributed only versus total neutron decay rate,
for various values of the GUT scale parameters, when the \co{13} operator is
included.}}
 
\vbox{$$\hskip -.3125 in
\begin{array}{|c|ccc|ccc|ccc|}
\hline
\rm{run\atop no.} & \multicolumn{3}{|c|}{\frac{\rx K^+ \overline\nu.^{\rm
gluino}}{\rx K^+ \overline\nu.^{\rm total}}} &
\multicolumn{3}{|c|}{\frac{\rx \pi^+ \overline\nu.^{\rm gluino}}{\rx \pi^+
\overline\nu.^{\rm total}}} &
\multicolumn{3}{|c|}{\frac{\rx K^0 \mu^+.^{\rm gluino}}{\rx K^0 \mu^+.^{\rm
total}}} \\
\cline{2-10} & \max & \beta=-\alpha & \min& \max & \beta=-\alpha & \min& \max &
\beta=-\alpha & \min \\
\hline
 \rm I & 0.019 & 0.010 & 0.010 & 0.0060 & 0.0038 & 0.0038 & 0.00038 & 0.00038 &
0.00038\cr
 \rm II & 0.085 & 0.053 & 0.051 & 0.036 & 0.024 & 0.023 & 0.00067 & 0.00067 &
0.00067\cr
 \rm III(1) & 0.25 & 0.15 & 0.11 & 0.035 & 0.030 & 0.023 & 0.020 & 0.020 &
0.020\cr
 \rm III(2) & 0.16 & 0.10 & 0.076 & 0.016 & 0.014 & 0.012 & 0.013 & 0.013 &
0.013\cr
 \rm III(3) & 0.10 & 0.071 & 0.055 & 0.0099 & 0.0090 & 0.0078 & 0.010 & 0.010 &
0.010\cr
\hline
\end{array}$$ {\small Table A3A: Ratios of the rate of proton decay that would
occur if gluino diagrams contributed only versus total proton decay rate for
the three most dominant decay modes, for various values of the GUT scale
parameters, when the \co{13} operator is included.}}

\protect
\vbox{$$ \hskip -1.125 in
\begin{array}{|c|ccc|ccc|ccc|ccc|}
\hline
\rm{run\atop no.} & 
\multicolumn{3}{|c|}{\frac{\rnx K^0 \overline\nu.^{\rm gluino}}{\rnx K^0
\overline\nu.^{\rm total}}} &
\multicolumn{3}{|c|}{\frac{\rnx \pi^0 \overline\nu.^{\rm gluino}}{\rnx \pi^0
\overline\nu.^{\rm total}}} &
\multicolumn{3}{|c|}{\frac{\rnx \eta \overline\nu.^{\rm gluino}}{\rnx \eta
\overline\nu.^{\rm total}}} &
\multicolumn{3}{|c|}{\frac{\rnx \pi^- \mu^+.^{\rm gluino}}{\rnx \pi^-
\mu^+.^{\rm total}}} \\
\cline{2-13} & \max & \beta=-\alpha & \min& \max & \beta=-\alpha & \min& \max &
\beta=-\alpha & \min & \max & \beta=-\alpha & \min\\
\hline
 \rm I & 0.0055 & 0.0034 & 0.0034 & 0.0060 & 0.0038 & 0.0038 & 0.025 & 0.025 &
0.019 & 0.0099 & 0.0099 & 0.0099\cr
 \rm II & 0.033 & 0.021 & 0.020 & 0.036 & 0.024 & 0.023 & 0.11 & 0.11 & 0.087 &
0.043 & 0.043 & 0.043\cr
 \rm III(1) & 0.033 & 0.028 & 0.023 & 0.035 & 0.030 & 0.023 & 0.83 & 0.41 &
0.32 & 0.20 & 0.20 & 0.20\cr
 \rm III(2) & 0.014 & 0.012 & 0.010 & 0.016 & 0.014 & 0.012 & 0.60 & 0.28 &
0.22 & 0.20 & 0.20 & 0.20\cr
 \rm III(3) & 0.0087 & 0.0080 & 0.0070 & 0.0099 & 0.0090 & 0.0078 & 0.47 & 0.22
& 0.17 & 0.16 & 0.16 & 0.16\cr
\hline
\end{array}$$ {\small Table A3B: Ratios of the rate of neutron decay that would
occur if gluino diagrams contributed only versus total neutron decay rate for
various values of the GUT scale parameters, when the \co{13} operator is
included.}}
 
\vbox{$$\begin{array}{|c|ccc|ccc|}
\hline
\rm {run \atop no.} & \multicolumn{3}{|c|}{\sqrt{\frac{\rx K^+
{\overline\nu_l}.^{\rm LLRR}} {\rx K^+ {\overline\nu_l}.^{\rm LLLL}}}} &
\multicolumn{3}{|c|}{\sqrt{\frac{\rx \pi^+ {\overline\nu_l}.^{\rm LLRR}} {\rx
\pi^+ {\overline\nu_l}.^{\rm LLLL}}}}\\
\cline{2-7} & \overline\nu_e & \overline\nu_\mu & \overline\nu_\tau &
\overline\nu_e & \overline\nu_\mu & \overline\nu_\tau  \\
\hline
 \rm I & 0.000084 & 0.011 & 2.0 & 0.00019 & 0.024 & 7.3\cr
 \rm II & 0.000086 & 0.011 & 1.8 & 0.00020 & 0.025 & 6.4\cr
 \rm III(1) & 0.00021 & 0.029 & 3.8 & 0.00056 & 0.068 & 10.\cr
 \rm III(2) & 0.00026 & 0.040 & 4.6 & 0.00067 & 0.097 & 14.\cr
 \rm III(3) & 0.00031 & 0.048 & 5.7 & 0.00080 & 0.12 & 17.\cr
\hline
\end{array}$$ {\small Table A4: Ratios of the rate of proton decay that would
occur if LLLL operators contributed only versus the rate of proton decay that
would occur if LLRR operators contributed only, for each of the three
anti-neutrino generations, for various values of the GUT scale parameters, when
the \co{13} operator is included.}}
\vbox{$$
\begin{array}{|c|ccccc|}
\hline
\rm{run\atop no.}& \sqrt{\frac{\rrx K^+\overline\nu_e.^{\rm LLLL\atop}}{\rrx
K^+\overline\nu_\tau.^{\rm LLRR\atop}}} &
\sqrt{\frac{\rrx K^+\overline\nu_e.^{\rm LLRR\atop}}{\rrx
K^+\overline\nu_\tau.^{\rm LLRR\atop}}} &
\sqrt{\frac{\rrx K^+\overline\nu_\mu.^{\rm LLLL\atop}}{\rrx
K^+\overline\nu_\tau.^{\rm LLRR\atop}}} &
\sqrt{\frac{\rrx K^+\overline\nu_\mu.^{\rm LLRR\atop}}{\rrx
K^+\overline\nu_\tau.^{\rm LLRR\atop}}} &
\sqrt{\frac{\rrx K^+\overline\nu_\tau.^{\rm LLLL\atop}}{\rrx
K^+\overline\nu_\tau.^{\rm LLRR\atop}}} \\
\hline
 \rm I & 0.082 & 6.8 \times 10^{-6} & 1.5 & 0.017 & 0.49\cr
 \rm II & 0.10 & 8.6 \times 10^{-6} & 1.9 & 0.021 & 0.56\cr
 \rm III(1) & 0.035 & 7.2 \times 10^{-6} & 0.60 & 0.017 & 0.26\cr
 \rm III(2) & 0.032 & 8.2 \times 10^{-6} & 0.49 & 0.020 & 0.22\cr
 \rm III(3) & 0.026 & 8.1 \times 10^{-6} & 0.41 & 0.020 & 0.18\cr
\hline
\end{array}$$ {\small Table A5A: Ratios of partial decay rates for $p\to K^+
\overline\nu$, which compare the importance of the LLLL and LLRR operators for
each generation of anti-neutrino versus contribution of the LLRR operator of
the third generation anti-neutrino for various values of the GUT scale
parameters, when the \co{13} operator is included.}} 

\vbox{$$
\begin{array}{|c|ccccc|}
\hline
\rm{run\atop no.}& 
\sqrt{\frac{\rrx \pi^+\overline\nu_e.^{\rm LLLL\atop}}{\rrx
\pi^+\overline\nu_\tau.^{\rm LLRR\atop}}} &
\sqrt{\frac{\rrx \pi^+\overline\nu_e.^{\rm LLRR\atop}}{\rrx
\pi^+\overline\nu_\tau.^{\rm LLRR\atop}}} &
\sqrt{\frac{\rrx \pi^+\overline\nu_\mu.^{\rm LLLL\atop}}{\rrx
\pi^+\overline\nu_\tau.^{\rm LLRR\atop}}} &
\sqrt{\frac{\rrx \pi^+\overline\nu_\mu.^{\rm LLRR\atop}}{\rrx
\pi^+\overline\nu_\tau.^{\rm LLRR\atop}}} &
\sqrt{\frac{\rrx \pi^+\overline\nu_\tau.^{\rm LLLL\atop}}{\rrx
\pi^+\overline\nu_\tau.^{\rm LLRR\atop}}} \\
\hline
 \rm I & 0.026 & 4.8 \times 10^{-6} & 0.48 & 0.012 & 0.14\cr
 \rm II & 0.030 & 6.1 \times 10^{-6} & 0.59 & 0.015 & 0.16\cr
 \rm III(1) & 0.0095 & 5.3 \times 10^{-6} & 0.18 & 0.013 & 0.098\cr
 \rm III(2) & 0.0081 & 5.4 \times 10^{-6} & 0.13 & 0.013 & 0.071\cr
 \rm III(3) & 0.0067 & 5.4 \times 10^{-6} & 0.11 & 0.013 & 0.058\cr
\hline
\end{array}$$ {\small Table A5B: Ratios of partial decay rates for $p\to\pi^+
\overline\nu$, which compare the importance of the LLLL and LLRR operators for
each generation of anti-neutrino versus contribution of the LLRR operator of
the third generation anti-neutrino for various values of the GUT scale
parameters, when the \co{13} operator is included.}}

\vbox{$$\hskip -.09375 in
\begin{array}{|c|rrrrr|}
\hline
\rm {\mathstrut run \atop no.\mathstrut}&
 \rm I  &   \rm II  &   \rm III(1)  &   \rm III(2)  &   \rm III(3)  \cr
\hline
\td\alpha_{GUT}^{-1} & 24.43  &  24.36  &  24.51  &  24.65  &  24.75  \cr
\mg & 2.498 \times {{10}^{16}}  &  3.172 \times {{10}^{16}}  &  
				3.327 \times {{10}^{16}}  &  2.857 \times {{10}^{16}}  &  
				2.513 \times {{10}^{16}}  \cr
\ep_3 & -0.04760  &  -0.04886  &  -0.04342  &  -0.04420  &  -0.04550  \cr
 \hline A & 0.7640  &  0.8067  &  0.8523  &  0.8867  &  0.8872  \cr B &
0.05259  &  0.05439  &  0.05630  &  0.05882  &  0.05956  \cr C & 0.0001096  & 
0.0001155  &  0.0001213  &  0.0001231  &  0.0001226  \cr E & 0.01251  & 
0.01308  &  0.01360  &  0.01397  &  0.01397  \cr
\phi & 1.066  &  1.041  &  1.020  &  1.023  &  1.038  \cr D & 0.0004633  & 
0.0004944  &  0.0005064  &  0.0005691  &  0.0005665  \cr
\delta & 5.698  &  5.706  &  5.698  &  5.742  &  5.744  \cr
\hline
\tan\beta & 52.77  &  54.38  &  55.39  &  55.86  &  55.92  \cr
\mu(M_Z) & 80.0  &  80.0  &  160.  &  240.  &  300.  \cr m_{1/2} & 280.  & 
240.  &  170.  &  170.  &  170.  \cr m_0 & 400.  &  700.  &  1400.  &  1400. 
&  1400.  \cr m_{H_d} & 706.4 & 994.6 & 1858. & 1859. & 1855.  \cr m_{H_u} &
635.6 & 865.3 & 1599. & 1591. & 1585. \cr A_0 & 322.2 &  458.4 &  -982.4 & 
-1079. &  -1274. \cr
\hline
\end{array}$$ {\small Table A6: Values of the GUT scale parameters used in
Tables A1 through A5. All dimensions in GeV units.}}

\normalsize
\protect
\vbox{\begin{center}
$$\begin{array}{|c|c|ccccc|}
\hline
\rm{run\atop no.} & \lower .1 in \hbox{${{{\rm m \atop \rm o}\atop \rm d}\atop
\rm e}\atop \rm
\scriptscriptstyle l$} & 
\frac{\tau (p \to K^+ \overline\nu)}{10^{32} {\rm yrs}} & \frac{\G(p \to \pi^+
\overline\nu)}{\G(p \to K^+ \overline\nu)} &
\frac{\rr K^0 \mu^+.}{\rr K^+ \overline\nu.} &
\frac{\rr \pi^0 \mu^+.}{\rr K^+ \overline\nu.} &
\frac{\rr \eta \mu^+.}{\rr K^+ \overline\nu.} \\
\hline & a & 7.3 & 0.22 & 0.0020 & 0.0011 & 0.00039\cr
 & b & 3.5 & 0.17 & 0.0032 & 0.0017 & 0.00060\cr
 \rm I & c & 11. & 0.25 & 0.0012 & 0.00067 & 0.00024\cr
 & d & 4.1 & 0.15 & 0.0047 & 0.0023 & 0.00082\cr
 & e & 14. & 0.25 & 0.0028 & 0.0013 & 0.00045\cr
 & f & 2.0 & 0.13 & 0.0049 & 0.0024 & 0.00086\cr
 \hline & a & 23. & 0.21 & 0.0028 & 0.0015 & 0.00054\cr
 & b & 10. & 0.16 & 0.0040 & 0.0021 & 0.00076\cr
 \rm II & c & 36. & 0.24 & 0.0018 & 0.00098 & 0.00034\cr
 & d & 10. & 0.15 & 0.0050 & 0.0025 & 0.00087\cr
 & e & 39. & 0.23 & 0.0036 & 0.0016 & 0.00055\cr
 & f & 4.9 & 0.13 & 0.0051 & 0.0026 & 0.00090\cr
 \hline & a & 59. & 0.30 & 0.00078 & 0.00034 & 0.00011\cr
 & b & 68. & 0.28 & 0.0023 & 0.0012 & 0.00039\cr
 \rm III(1) & c & 68. & 0.31 & 0.00052 & 0.00017 & 0.000044\cr
 & d & 32. & 0.24 & 0.0016 & 0.00072 & 0.00025\cr
 & e & 57. & 0.30 & 0.00080 & 0.00025 & 0.000076\cr
 & f & 21. & 0.21 & 0.0021 & 0.00097 & 0.00034\cr
 \hline & a & 26. & 0.31 & 0.00048 & 0.00017 & 0.000048\cr
 & b & 33. & 0.32 & 0.0013 & 0.00059 & 0.00019\cr
 \rm III(2) & c & 29. & 0.32 & 0.00039 & 0.000094 & 0.000019\cr
 & d & 18. & 0.27 & 0.0010 & 0.00042 & 0.00014\cr
 & e & 26. & 0.32 & 0.00053 & 0.00014 & 0.000036\cr
 & f & 13. & 0.24 & 0.0014 & 0.00062 & 0.00021\cr
 \hline & a & 18. & 0.31 & 0.00040 & 0.00012 & 0.000031\cr
 & b & 23. & 0.34 & 0.0010 & 0.00040 & 0.00012\cr
 \rm III(3) & c & 20. & 0.32 & 0.00034 & 0.000073 & 0.000012\cr
 & d & 14. & 0.28 & 0.00081 & 0.00031 & 0.000099\cr
 & e & 19. & 0.32 & 0.00044 & 0.00010 & 0.000024\cr
 & f & 11. & 0.26 & 0.0011 & 0.00047 & 0.00016\cr
\hline
\end{array}$$
\end{center} {\small Table B1A: Partial mean lifetime for proton decaying into
kaon plus anti-neutrino and ratios of the rates of proton decay into various
decay products versus rate of decay into kaon plus anti-neutrino for various
values of the GUT scale parameters, when the \co{13} operator is not included.
For all entries, $\beta=-\alpha.$}}

\vbox{\begin{center}
$$
\begin{array}{|c|c|cccc|}
\hline
\rm{run\atop no.} & \lower .1 in \hbox{${{{\rm m \atop \rm o}\atop \rm d}\atop
\rm e}\atop \rm
\scriptscriptstyle l$} & \frac{\tau (n \to K^0 \overline\nu)}{10^{32} {\rm
yrs}}  & \frac{\G(n \to \pi^0 \overline\nu)}{\G(n \to K^0 \overline\nu)} &
\frac{\rn \eta\overline\nu.}{\rn K^0 \overline\nu.} &
\frac{\rn \pi^- \mu^+.}{\rn K^0 \overline\nu.} \\
\hline 
 & a & 2.4 & 0.036 & 0.0020 & 0.00071\cr
 & b & 1.4 & 0.033 & 0.0047 & 0.0013\cr
 \rm I
 & c & 3.1 & 0.036 & 0.00095 & 0.00038\cr
 & d & 1.9 & 0.035 & 0.010 & 0.0022\cr
 & e & 4.0 & 0.036 & 0.0039 & 0.00073\cr
 & f & 1.0 & 0.035 & 0.012 & 0.0025\cr
 \hline
 & a & 8.0 & 0.035 & 0.0028 & 0.0011\cr
 & b & 4.2 & 0.033 & 0.0063 & 0.0018\cr
 \rm II
 & c & 11. & 0.036 & 0.0015 & 0.00059\cr
 & d & 4.8 & 0.035 & 0.010 & 0.0024\cr
 & e & 12. & 0.036 & 0.0044 & 0.0010\cr
 & f & 2.5 & 0.034 & 0.011 & 0.0026\cr
 \hline
 & a & 16. & 0.039 & 0.0011 & 0.00018\cr
 & b & 17. & 0.036 & 0.0038 & 0.00060\cr
 \rm III(1)
 & c & 17. & 0.038 & 0.00075 & 0.000082\cr
 & d & 9.9 & 0.037 & 0.0015 & 0.00044\cr
 & e & 14. & 0.036 & 0.00037 & 0.00012\cr
 & f & 7.4 & 0.036 & 0.0026 & 0.00068\cr
 \hline
 & a & 6.6 & 0.039 & 0.00066 & 0.000086\cr
 & b & 7.4 & 0.037 & 0.0020 & 0.00027\cr
 \rm III(2)
 & c & 7.0 & 0.038 & 0.00051 & 0.000045\cr
 & d & 5.0 & 0.037 & 0.00085 & 0.00023\cr
 & e & 6.1 & 0.036 & 0.00028 & 0.000064\cr
 & f & 4.1 & 0.037 & 0.0015 & 0.00038\cr
 \hline
 & a & 4.5 & 0.039 & 0.00054 & 0.000060\cr
 & b & 5.1 & 0.037 & 0.0014 & 0.00018\cr
 \rm III(3)
 & c & 4.8 & 0.038 & 0.00044 & 0.000034\cr
 & d & 3.7 & 0.037 & 0.00064 & 0.00017\cr
 & e & 4.3 & 0.037 & 0.00026 & 0.000047\cr
 & f & 3.2 & 0.038 & 0.0011 & 0.00028\cr
\hline
\end{array}$$
\end{center} {\small Table B1B: Partial mean lifetime for neutron decaying into
kaon plus anti-neutrino and ratios of the rates of neutron decay into various
decay products versus rate of decay into kaon plus anti-neutrino for various
values of the GUT scale parameters, when the \co{13} operator is not included.
For all entries, $\beta=-\alpha.$}}

\normalsize
\vbox{$$\hskip -.125 in
\begin{array}{|c|c|ccc|ccc|ccc|}
\hline
\rm{run\atop no.}& \lower .1 in \hbox{${{{\rm m \atop \rm o}\atop \rm d}\atop
\rm e}\atop \rm
\scriptscriptstyle l$} &
 \multicolumn{3}{|c|}{\frac{\rx K^+
\overline\nu.^{\rm chargino}}{\rx K^+ \overline\nu.^{\rm total}}} &
\multicolumn{3}{|c|}{\frac{\rx \pi^+ \overline\nu.^{\rm chargino}}{\rx \pi^+
\overline\nu.^{\rm total}}} &
\multicolumn{3}{|c|}{\frac{\rx K^0 \mu^+.^{\rm chargino}}{\rx K^0 \mu^+.^{\rm
total}}} \\
\cline{3-11}  && \max & \beta=-\alpha & \min& \max & \beta=-\alpha & \min&
\max &
\beta=-\alpha & \min \\
\hline & a & 1.2 & 1.1 & 0.85 & 1.1 & 1.1 & 0.87 & 1.0 & 1.0 & 1.0\cr
 & b & 1.2 & 1.2 & 0.93 & 1.2 & 1.2 & 0.86 & 1.0 & 1.0 & 1.0\cr
 \rm I & c & 1.2 & 1.1 & 0.83 & 1.1 & 1.1 & 0.89 & 0.99 & 0.99 & 0.99\cr
 & d & 1.2 & 0.99 & 0.98 & 1.2 & 0.92 & 0.91 & 1.0 & 1.0 & 1.0\cr
 & e & 1.2 & 0.91 & 0.91 & 1.1 & 0.91 & 0.91 & 1.0 & 1.0 & 1.0\cr
 & f & 1.2 & 1.0 & 1.0 & 1.2 & 0.97 & 0.94 & 1.0 & 1.0 & 1.0\cr
 \hline & a & 1.5 & 1.3 & 0.72 & 1.4 & 1.3 & 0.72 & 1.0 & 1.0 & 1.0\cr
 & b & 1.4 & 1.4 & 0.90 & 1.4 & 1.4 & 0.75 & 1.0 & 1.0 & 1.0\cr
 \rm II & c & 1.4 & 1.3 & 0.65 & 1.3 & 1.2 & 0.75 & 1.0 & 1.0 & 1.0\cr
 & d & 1.4 & 1.0 & 1.0 & 1.4 & 0.88 & 0.86 & 1.0 & 1.0 & 1.0\cr
 & e & 1.4 & 0.84 & 0.84 & 1.3 & 0.81 & 0.81 & 1.0 & 1.0 & 1.0\cr
 & f & 1.4 & 1.1 & 1.1 & 1.4 & 0.97 & 0.93 & 1.0 & 1.0 & 1.0\cr
 \hline & a & 1.9 & 1.5 & 0.53 & 1.5 & 1.3 & 0.69 & 1.1 & 1.1 & 1.1\cr
 & b & 2.6 & 2.4 & 0.34 & 2.0 & 2.0 & 0.47 & 1.2 & 1.2 & 1.2\cr
 \rm III(1) & c & 1.6 & 1.4 & 0.62 & 1.3 & 1.2 & 0.76 & 1.2 & 1.2 & 1.2\cr
 & d & 2.7 & 0.48 & 0.42 & 2.0 & 0.53 & 0.50 & 1.1 & 1.1 & 1.1\cr
 & e & 1.7 & 0.65 & 0.65 & 1.3 & 0.76 & 0.76 & 0.98 & 0.98 & 0.98\cr
 & f & 2.8 & 0.53 & 0.42 & 2.3 & 0.50 & 0.42 & 1.1 & 1.1 & 1.1\cr
 \hline & a & 1.5 & 1.3 & 0.65 & 1.3 & 1.2 & 0.78 & 1.1 & 1.1 & 1.1\cr
 & b & 2.2 & 2.0 & 0.42 & 1.6 & 1.6 & 0.59 & 1.2 & 1.2 & 1.2\cr
 \rm III(2) & c & 1.4 & 1.2 & 0.72 & 1.2 & 1.2 & 0.83 & 1.1 & 1.1 & 1.1\cr
 & d & 2.2 & 0.52 & 0.47 & 1.6 & 0.62 & 0.60 & 1.0 & 1.0 & 1.0\cr
 & e & 1.4 & 0.74 & 0.74 & 1.2 & 0.83 & 0.83 & 0.99 & 0.99 & 0.99\cr
 & f & 2.5 & 0.51 & 0.41 & 1.9 & 0.56 & 0.50 & 1.1 & 1.1 & 1.1\cr
 \hline & a & 1.4 & 1.3 & 0.70 & 1.2 & 1.2 & 0.81 & 1.1 & 1.1 & 1.1\cr
 & b & 1.9 & 1.8 & 0.48 & 1.5 & 1.5 & 0.65 & 1.1 & 1.1 & 1.1\cr
 \rm III(3) & c & 1.3 & 1.2 & 0.77 & 1.2 & 1.1 & 0.86 & 1.1 & 1.1 & 1.1\cr
 & d & 1.9 & 0.56 & 0.52 & 1.5 & 0.67 & 0.66 & 1.0 & 1.0 & 1.0\cr
 & e & 1.3 & 0.78 & 0.78 & 1.2 & 0.86 & 0.86 & 0.99 & 0.99 & 0.99\cr
 & f & 2.3 & 0.53 & 0.44 & 1.7 & 0.61 & 0.56 & 1.1 & 1.1 & 1.1\cr
\hline
\end{array}$$ {\small Table B2A: Ratios of the rate of proton decay that would
occur if chargino diagrams contributed only versus total proton decay rate for
the three most dominant decay modes, for various values of the GUT scale
parameters, when the \co{13} operator is not included.}}

\vbox{$$\hskip -.09375 in
\begin{array}{|c|c|ccc|ccc|ccc|}
\hline
\rm{run\atop no.} & \lower .1 in \hbox{${{{\rm m \atop \rm o}\atop \rm d}\atop
\rm e}\atop \rm
\scriptscriptstyle l$} &
\multicolumn{3}{|c|}{\frac{\rnx K^0 \overline\nu.^{\rm chargino}}{\rnx K^0
\overline\nu.^{\rm total}}} &
\multicolumn{3}{|c|}{\frac{\rnx \pi^0 \overline\nu.^{\rm chargino}}{\rnx \pi^0
\overline\nu.^{\rm total}}} &
\multicolumn{3}{|c|}{\frac{\rnx \eta \overline\nu.^{\rm chargino}}{\rnx \eta
\overline\nu.^{\rm total}}} \\
\cline{3-11} && \max & \beta=-\alpha & \min& \max & \beta=-\alpha & \min& \max
&
\beta=-\alpha & \min \\
\hline 
 & a & 1.2 & 1.1 & 0.85 & 1.1 & 1.1 & 0.87 & 1.2 & 1.1 & 1.0 \cr
 & b & 1.2 & 1.2 & 0.86 & 1.2 & 1.2 & 0.86 & 1.1 & 1.1 & 1.1 \cr
 \rm I & c & 1.1 & 1.1 & 0.87 & 1.1 & 1.1 & 0.89 & 1.2 & 0.95 & 0.92 \cr
 & d & 1.2 & 0.92 & 0.92 & 1.2 & 0.92 & 0.91 & 1.1 & 1.1 & 1.1 \cr
 & e & 1.1 & 0.91 & 0.91 & 1.1 & 0.91 & 0.91 & 1.2 & 1.2 & 1.1 \cr
 & f & 1.2 & 0.98 & 0.96 & 1.2 & 0.97 & 0.94 & 1.1 & 1.1 & 1.1 \cr
 \hline & a & 1.4 & 1.3 & 0.70 & 1.4 & 1.3 & 0.72 & 1.4 & 1.1 & 1.1 \cr
 & b & 1.4 & 1.4 & 0.76 & 1.4 & 1.4 & 0.75 & 1.3 & 1.1 & 1.1 \cr
 \rm II & c & 1.3 & 1.3 & 0.71 & 1.3 & 1.2 & 0.75 & 1.4 & 0.92 & 0.88 \cr
 & d & 1.4 & 0.89 & 0.88 & 1.4 & 0.88 & 0.86 & 1.3 & 1.3 & 1.2 \cr
 & e & 1.3 & 0.81 & 0.81 & 1.3 & 0.81 & 0.81 & 1.4 & 1.4 & 1.2 \cr
 & f & 1.4 & 0.99 & 0.97 & 1.4 & 0.97 & 0.93 & 1.3 & 1.3 & 1.2 \cr
 \hline & a & 1.6 & 1.4 & 0.64 & 1.5 & 1.3 & 0.69 & 3.1 & 0.53 & 0.43 \cr
 & b & 2.1 & 2.1 & 0.43 & 2.0 & 2.0 & 0.47 & 1.9 & 0.64 & 0.63 \cr
 \rm III(1) & c & 1.4 & 1.3 & 0.72 & 1.3 & 1.2 & 0.76 & 3.1 & 0.35 & 0.31 \cr
 & d & 2.2 & 0.49 & 0.47 & 2.0 & 0.53 & 0.50 & 2.1 & 2.1 & 0.87 \cr
 & e & 1.4 & 0.75 & 0.75 & 1.3 & 0.76 & 0.76 & 3.1 & 2.9 & 0.65 \cr
 & f & 2.5 & 0.46 & 0.41 & 2.3 & 0.50 & 0.42 & 1.8 & 1.8 & 0.94 \cr
 \hline & a & 1.3 & 1.3 & 0.74 & 1.3 & 1.2 & 0.78 & 3.1 & 0.48 & 0.38 \cr
 & b & 1.7 & 1.7 & 0.55 & 1.6 & 1.6 & 0.59 & 2.2 & 0.52 & 0.51 \cr
 \rm III(2) & c & 1.2 & 1.2 & 0.81 & 1.2 & 1.2 & 0.83 & 2.8 & 0.40 & 0.36 \cr
 & d & 1.7 & 0.58 & 0.57 & 1.6 & 0.62 & 0.60 & 2.4 & 2.4 & 0.72 \cr
 & e & 1.2 & 0.83 & 0.82 & 1.2 & 0.83 & 0.83 & 2.7 & 2.6 & 0.55 \cr
 & f & 2.1 & 0.52 & 0.47 & 1.9 & 0.56 & 0.50 & 2.1 & 2.0 & 0.82 \cr
 \hline & a & 1.3 & 1.2 & 0.78 & 1.2 & 1.2 & 0.81 & 2.9 & 0.49 & 0.39 \cr
 & b & 1.6 & 1.6 & 0.61 & 1.5 & 1.5 & 0.65 & 2.4 & 0.47 & 0.46 \cr
 \rm III(3) & c & 1.2 & 1.2 & 0.84 & 1.2 & 1.1 & 0.86 & 2.4 & 0.45 & 0.41 \cr
 & d & 1.6 & 0.64 & 0.63 & 1.5 & 0.67 & 0.66 & 2.5 & 2.5 & 0.64 \cr
 & e & 1.2 & 0.86 & 0.85 & 1.2 & 0.86 & 0.86 & 2.4 & 2.3 & 0.54 \cr
 & f & 1.9 & 0.57 & 0.53 & 1.7 & 0.61 & 0.56 & 2.2 & 2.1 & 0.75 \cr
\hline
\end{array}$$ {\small Table B2B: Ratios of the rate of neutron decay that would
occur if chargino diagrams contributed only versus total neutron decay rate for
various values of the GUT scale parameters, when the \co{13} operator is not
included.}}
 
\vbox{$$\begin{array}{|c|c|ccc|ccc|}
\hline
\rm {run \atop no.} & \lower .1 in \hbox{${{{\rm m \atop \rm o}\atop \rm
d}\atop
\rm e}\atop \rm
\scriptscriptstyle l$} &
 \multicolumn{3}{|c|}{\sqrt{\frac{\rx K^+ {\overline\nu_l}.^{\rm LLRR}} {\rx
K^+ {\overline\nu_l}.^{\rm LLLL}}}} &
\multicolumn{3}{|c|}{\sqrt{\frac{\rx \pi^+ {\overline\nu_l}.^{\rm LLRR}} {\rx
\pi^+ {\overline\nu_l}.^{\rm LLLL}}}}\\
\cline{3-8}  && \overline\nu_e & \overline\nu_\mu & \overline\nu_\tau &
\overline\nu_e & \overline\nu_\mu & \overline\nu_\tau  \\
\hline 
 & a & 0.000044 & 0.011 & 4.2 & 0.000073 & 0.018 & 8.4\cr
 & b & 0.000022 & 0.0049 & 1.9 & 0.000037 & 0.0083 & 3.6\cr
 \rm I & c & 0.000065 & 0.017 & 5.7 & 0.00011 & 0.029 & 12.\cr
 & d & 0.000022 & 0.0043 & 1.7 & 0.000037 & 0.0071 & 2.8\cr
 & e & 0.000062 & 0.011 & 4.2 & 0.00010 & 0.018 & 6.5\cr
 & f & 0.000015 & 0.0031 & 1.3 & 0.000025 & 0.0050 & 2.1\cr
 \hline & a & 0.000048 & 0.011 & 5.1 & 0.000078 & 0.018 & 10.\cr
 & b & 0.000023 & 0.0049 & 2.6 & 0.000038 & 0.0083 & 5.4\cr
 \rm II & c & 0.000071 & 0.017 & 5.7 & 0.00012 & 0.029 & 11.\cr
 & d & 0.000022 & 0.0044 & 2.2 & 0.000037 & 0.0073 & 3.5\cr
 & e & 0.000061 & 0.011 & 4.2 & 0.00010 & 0.019 & 6.4\cr
 & f & 0.000015 & 0.0031 & 1.7 & 0.000025 & 0.0051 & 2.7\cr
 \hline & a & 0.00019 & 0.032 & 6.4 & 0.00031 & 0.055 & 10.\cr
 & b & 0.000071 & 0.013 & 3.4 & 0.00012 & 0.023 & 5.8\cr
 \rm III(1) & c & 0.00030 & 0.045 & 7.6 & 0.00049 & 0.079 & 12.\cr
 & d & 0.000059 & 0.014 & 3.8 & 0.000098 & 0.022 & 6.8\cr
 & e & 0.00015 & 0.042 & 8.5 & 0.00025 & 0.064 & 16.\cr
 & f & 0.000042 & 0.0093 & 2.8 & 0.000069 & 0.015 & 4.8\cr
 \hline & a & 0.00029 & 0.048 & 10. & 0.00046 & 0.081 & 16.\cr
 & b & 0.00011 & 0.020 & 5.3 & 0.00018 & 0.034 & 9.1\cr
 \rm III(2) & c & 0.00044 & 0.066 & 12. & 0.00072 & 0.12 & 19.\cr
 & d & 0.000088 & 0.020 & 6.1 & 0.00014 & 0.032 & 11.\cr
 & e & 0.00023 & 0.061 & 13. & 0.00037 & 0.095 & 24.\cr
 & f & 0.000062 & 0.014 & 4.4 & 0.00010 & 0.022 & 7.6\cr
 \hline & a & 0.00035 & 0.058 & 13. & 0.00056 & 0.099 & 20.\cr
 & b & 0.00013 & 0.024 & 6.8 & 0.00022 & 0.042 & 12.\cr
 \rm III(3) & c & 0.00054 & 0.081 & 15. & 0.00088 & 0.14 & 24.\cr
 & d & 0.00011 & 0.024 & 7.8 & 0.00018 & 0.040 & 14.\cr
 & e & 0.00028 & 0.075 & 16. & 0.00046 & 0.12 & 30.\cr
 & f & 0.000076 & 0.017 & 5.6 & 0.00012 & 0.027 & 9.7\cr
\hline
\end{array}$$ {\small Table B4: Ratios of the rate of proton decay that would
occur if LLLL operators contributed only versus the rate of proton decay that
would occur if LLRR operators contributed only, for each of the three
anti-neutrino generations, for various values of the GUT scale parameters, when
the \co{13} operator is not included.}}

\vbox{$$
\begin{array}{|c|c|ccccc|}
\hline
\rm{run\atop no.} & \lower .1 in \hbox{${{{\rm m \atop \rm o}\atop \rm d}\atop
\rm e}\atop \rm
\scriptscriptstyle l$} &
 \sqrt{\frac{\rrx K^+\overline\nu_e.^{\rm LLLL\atop}}{\rrx
K^+\overline\nu_\tau.^{\rm LLRR\atop}}} &
\sqrt{\frac{\rrx K^+\overline\nu_e.^{\rm LLRR\atop}}{\rrx
K^+\overline\nu_\tau.^{\rm LLRR\atop}}} &
\sqrt{\frac{\rrx K^+\overline\nu_\mu.^{\rm LLLL\atop}}{\rrx
K^+\overline\nu_\tau.^{\rm LLRR\atop}}} &
\sqrt{\frac{\rrx K^+\overline\nu_\mu.^{\rm LLRR\atop}}{\rrx
K^+\overline\nu_\tau.^{\rm LLRR\atop}}} &
\sqrt{\frac{\rrx K^+\overline\nu_\tau.^{\rm LLLL\atop}}{\rrx
K^+\overline\nu_\tau.^{\rm LLRR\atop}}}  \\
\hline 
 & a & 0.062 & 2.8 \times 10^{-6} & 0.78 & 0.0084 & 0.24\cr
 & b & 0.13 & 2.8 \times 10^{-6} & 1.7 & 0.0085 & 0.53\cr
 \rm I & c & 0.043 & 2.8 \times 10^{-6} & 0.49 & 0.0085 & 0.18\cr
 & d & 0.12 & 2.8 \times 10^{-6} & 1.9 & 0.0084 & 0.57\cr
 & e & 0.045 & 2.8 \times 10^{-6} & 0.80 & 0.0085 & 0.24\cr
 & f & 0.17 & 2.7 \times 10^{-6} & 2.7 & 0.0082 & 0.79\cr
 \hline & a & 0.071 & 3.4 \times 10^{-6} & 0.95 & 0.010 & 0.20\cr
 & b & 0.15 & 3.5 \times 10^{-6} & 2.1 & 0.011 & 0.38\cr
 \rm II & c & 0.049 & 3.5 \times 10^{-6} & 0.61 & 0.010 & 0.18\cr
 & d & 0.15 & 3.4 \times 10^{-6} & 2.3 & 0.010 & 0.45\cr
 & e & 0.057 & 3.5 \times 10^{-6} & 0.94 & 0.011 & 0.24\cr
 & f & 0.22 & 3.3 \times 10^{-6} & 3.2 & 0.010 & 0.60\cr
 \hline & a & 0.015 & 3.0 \times 10^{-6} & 0.28 & 0.0091 & 0.16\cr
 & b & 0.043 & 3.0 \times 10^{-6} & 0.69 & 0.0092 & 0.30\cr
 \rm III(1) & c & 0.010 & 3.0 \times 10^{-6} & 0.21 & 0.0092 & 0.13\cr
 & d & 0.050 & 3.0 \times 10^{-6} & 0.67 & 0.0091 & 0.26\cr
 & e & 0.020 & 3.0 \times 10^{-6} & 0.22 & 0.0092 & 0.12\cr
 & f & 0.069 & 2.9 \times 10^{-6} & 0.95 & 0.0089 & 0.36\cr
 \hline & a & 0.010 & 3.0 \times 10^{-6} & 0.19 & 0.0090 & 0.10\cr
 & b & 0.029 & 3.0 \times 10^{-6} & 0.46 & 0.0091 & 0.19\cr
 \rm III(2) & c & 0.0068 & 3.0 \times 10^{-6} & 0.14 & 0.0091 & 0.085\cr
 & d & 0.034 & 3.0 \times 10^{-6} & 0.45 & 0.0090 & 0.16\cr
 & e & 0.013 & 3.0 \times 10^{-6} & 0.15 & 0.0091 & 0.076\cr
 & f & 0.047 & 2.9 \times 10^{-6} & 0.64 & 0.0088 & 0.23\cr
 \hline & a & 0.0086 & 3.0 \times 10^{-6} & 0.16 & 0.0091 & 0.080\cr
 & b & 0.023 & 3.0 \times 10^{-6} & 0.38 & 0.0092 & 0.15\cr
 \rm III(3) & c & 0.0056 & 3.0 \times 10^{-6} & 0.11 & 0.0091 & 0.068\cr
 & d & 0.028 & 3.0 \times 10^{-6} & 0.37 & 0.0090 & 0.13\cr
 & e & 0.011 & 3.0 \times 10^{-6} & 0.12 & 0.0092 & 0.061\cr
 & f & 0.038 & 2.9 \times 10^{-6} & 0.53 & 0.0089 & 0.18\cr
\hline
\end{array}$$ {\small Table B5A: Ratios of partial decay rates for $p\to K^+
\overline\nu$, which compare the importance of the LLLL and LLRR operators for
each generation of anti-neutrino versus contribution of the LLRR operator of
the third generation anti-neutrino for various values of the GUT scale
parameters, when the \co{13} operator is not included.}}

\vbox{$$
\begin{array}{|c|c|ccccc|}
\hline
\rm{run\atop no.} & \lower .1 in \hbox{${{{\rm m \atop \rm o}\atop \rm d}\atop
\rm e}\atop \rm
\scriptscriptstyle l$} &
\sqrt{\frac{\rrx \pi^+\overline\nu_e.^{\rm LLLL\atop}}{\rrx
\pi^+\overline\nu_\tau.^{\rm LLRR\atop}}} &
\sqrt{\frac{\rrx \pi^+\overline\nu_e.^{\rm LLRR\atop}}{\rrx
\pi^+\overline\nu_\tau.^{\rm LLRR\atop}}} &
\sqrt{\frac{\rrx \pi^+\overline\nu_\mu.^{\rm LLLL\atop}}{\rrx
\pi^+\overline\nu_\tau.^{\rm LLRR\atop}}} &
\sqrt{\frac{\rrx \pi^+\overline\nu_\mu.^{\rm LLRR\atop}}{\rrx
\pi^+\overline\nu_\tau.^{\rm LLRR\atop}}} &
\sqrt{\frac{\rrx \pi^+\overline\nu_\tau.^{\rm LLLL\atop}}{\rrx
\pi^+\overline\nu_\tau.^{\rm LLRR\atop}}} \\
\hline 
 & a & 0.038 & 2.8 \times 10^{-6} & 0.48 & 0.0085 & 0.12\cr
 & b & 0.076 & 2.8 \times 10^{-6} & 1.0 & 0.0085 & 0.28\cr
 \rm I & c & 0.026 & 2.8 \times 10^{-6} & 0.30 & 0.0085 & 0.084\cr
 & d & 0.076 & 2.8 \times 10^{-6} & 1.2 & 0.0085 & 0.35\cr
 & e & 0.027 & 2.8 \times 10^{-6} & 0.48 & 0.0085 & 0.15\cr
 & f & 0.11 & 2.7 \times 10^{-6} & 1.7 & 0.0083 & 0.48\cr
 \hline & a & 0.044 & 3.5 \times 10^{-6} & 0.58 & 0.010 & 0.095\cr
 & b & 0.092 & 3.5 \times 10^{-6} & 1.3 & 0.011 & 0.19\cr
 \rm II & c & 0.030 & 3.5 \times 10^{-6} & 0.37 & 0.011 & 0.088\cr
 & d & 0.095 & 3.5 \times 10^{-6} & 1.4 & 0.010 & 0.29\cr
 & e & 0.034 & 3.5 \times 10^{-6} & 0.57 & 0.011 & 0.16\cr
 & f & 0.13 & 3.4 \times 10^{-6} & 2.0 & 0.010 & 0.38\cr
 \hline & a & 0.0097 & 3.0 \times 10^{-6} & 0.17 & 0.0092 & 0.096\cr
 & b & 0.025 & 3.0 \times 10^{-6} & 0.40 & 0.0092 & 0.17\cr
 \rm III(1) & c & 0.0063 & 3.0 \times 10^{-6} & 0.12 & 0.0092 & 0.080\cr
 & d & 0.031 & 3.0 \times 10^{-6} & 0.42 & 0.0092 & 0.15\cr
 & e & 0.012 & 3.1 \times 10^{-6} & 0.14 & 0.0093 & 0.064\cr
 & f & 0.043 & 3.0 \times 10^{-6} & 0.60 & 0.0090 & 0.21\cr
 \hline & a & 0.0065 & 3.0 \times 10^{-6} & 0.11 & 0.0091 & 0.062\cr
 & b & 0.017 & 3.0 \times 10^{-6} & 0.27 & 0.0091 & 0.11\cr
 \rm III(2) & c & 0.0042 & 3.0 \times 10^{-6} & 0.078 & 0.0091 & 0.052\cr
 & d & 0.021 & 3.0 \times 10^{-6} & 0.28 & 0.0091 & 0.093\cr
 & e & 0.0081 & 3.0 \times 10^{-6} & 0.096 & 0.0092 & 0.042\cr
 & f & 0.029 & 2.9 \times 10^{-6} & 0.40 & 0.0089 & 0.13\cr
 \hline & a & 0.0054 & 3.0 \times 10^{-6} & 0.092 & 0.0091 & 0.049\cr
 & b & 0.014 & 3.0 \times 10^{-6} & 0.22 & 0.0092 & 0.085\cr
 \rm III(3) & c & 0.0034 & 3.0 \times 10^{-6} & 0.064 & 0.0092 & 0.042\cr
 & d & 0.017 & 3.0 \times 10^{-6} & 0.23 & 0.0091 & 0.072\cr
 & e & 0.0066 & 3.0 \times 10^{-6} & 0.079 & 0.0092 & 0.034\cr
 & f & 0.024 & 3.0 \times 10^{-6} & 0.33 & 0.0090 & 0.10\cr
\hline
\end{array}$$ {\small Table B5B: Ratios of partial decay rates for $p\to \pi^+
\overline\nu$, which compare the importance of the LLLL and LLRR operators for
each generation of anti-neutrino versus contribution of the LLRR operator of
the third generation anti-neutrino for various values of the GUT scale
parameters, when the \co{13} operator is not included.}}
 
\vbox{$$\hskip -.09375 in
\begin{array}{|c|rrrrr|}
\hline
\rm {\mathstrut run \atop no.\mathstrut}&
\rm I  &   \rm II  &   \rm III(1)  &   \rm III(2)  &   \rm III(3)  \cr
\hline
\td\alpha_{GUT}^{-1} & 24.43  &  24.36  &  24.51  &  24.65  &  24.75  \cr
\mg & 2.498 \times {{10}^{16}}  &  3.172 \times {{10}^{16}}  &  
				3.327 \times {{10}^{16}}  &  2.857 \times {{10}^{16}}  &
		  2.513 \times {{10}^{16}}  \cr
\ep_3 & -0.04760  &  -0.04886  &  -0.04342  &  -0.04420  &  -0.04550  \cr
\hline A & 0.7640  &  0.8067  &  0.8523  &  0.8867  &  0.8872  \cr B & 0.05798 
&  0.06019  &  0.06254  &  0.06533  &  0.06607  \cr C & 0.00008824  & 
0.00009204  &  0.00009550  &  0.00009801  &  0.00009809  \cr E & 0.01063  & 
0.01111  &  0.01154  &  0.01182  &  0.01180  \cr
\phi & 1.762  &  1.765  &  1.767  &  1.765  &  1.763  \cr
\hline
\tan\beta & 52.71  &  54.31  &  55.32  &  55.79  &  55.87  \cr
\mu(M_Z) & 80.0  &  80.0  &  160.  &  240.  &  300.  \cr m_{1/2} & 280.  & 
240.  &  170.  &  170.  &  170.  \cr m_0 & 400.  &  700.  &  1400.  &  1400. 
&  1400.  \cr m_{H_d} & 706.3 & 994.4 & 1858. & 1859. & 1855.  \cr m_{H_u} &
635.9 & 865.6 & 1599. & 1592. & 1585.  \cr A_0 & 322.2 &  458.4 &  -982.4 & 
-1079. &  -1274. \cr
\hline
\end{array}$$ {\small Table B6: Values of the GUT scale parameters used in
Tables B1 through B5. All dimensions in GeV units.}
\vskip 6 pt
\hskip -.75 in
\vrule height 1pt width 5.75in
\vskip -9pt
\hskip -.75 in
\vrule height 1pt width 5.75in}

\normalsize\def\G{\Gamma^{\vphantom{*}}}
\section{Discussion of the results}
\subsection{Overall Rates} Many significant results appear from the tables of
the previous section. First, comparing the rates of proton decay predicted in
the tables above with the results of experimental searches for proton and
baryon-number violating neutron decay summarized in Table \ref{t:experiment},
it can be seen that the predicted upper bounds on the lifetimes for nucleon
decay are above, and, in most cases, well above, the experimental lower
bounds. The loop integral
$I(M_{gaugino},M_{squark\atop (slepton)},M_{squark\atop (slepton)})$ goes
roughly like $1/M_{squark\atop (slepton)}^2$ in the limit where squarks and
sleptons are much heavier than gauginos. Hence, the decay rates go naively like
$(m_{1/2}^2+\mu_R^2)/m_0^4$, where $\mu_R \equiv \mu(M_Z)$. This approximation
roughly explains the dependence of the nucleon decay rates on $m_0$,
$\mu(M_Z)$, and $m_{1/2}$ seen in Tables A1A, A1B, B1A, and B1B.

\protect
\begin{table}
\vrule height 1pt width 4.75in
\vskip -10pt
\vrule height 1pt width 4.75in
\caption{Current experimental lower bounds on the various partial lifetimes of
the nucleons \protect\cite{pdg}}
\label{t:experiment}
$$\begin{array}{rcl}
\tau(p \to K^+ \overline\nu)&>& 1.0 \times 10^{32}\ \rm yrs\\
\tau(p \to \pi^+ \overline\nu)&>& .25 \times 10^{32}\ \rm yrs\\
\tau(p \to \pi^0 \mu^+)&>& 2.7 \times 10^{32}\ \rm yrs\\
\tau(p \to \eta \mu^+)&>& .69 \times 10^{32}\ \rm yrs\\
\tau(p \to \pi^0 e^+)&>& 5.5 \times 10^{32}\ \rm yrs\\
\tau(p \to \eta e^+)&>& 1.4 \times 10^{32}\ \rm yrs\\
\tau(n \to K^0 \overline\nu)&>& .86 \times 10^{32}\ \rm yrs\\
\tau(n \to \pi^0 \overline\nu)&>& 1.0 \times 10^{32}\ \rm yrs\\
\tau(n \to \eta\overline\nu)&>& .54 \times 10^{32}\ \rm yrs\\
\tau(n \to \pi^- \mu^+)&>& 1.0 \times 10^{32}\ \rm yrs\\
\tau(n \to \pi^- e^+)&>& 1.3 \times 10^{32}\ \rm yrs\\
\end{array}$$
\vrule height 1pt width 4.75in
\vskip -10pt
\vrule height 1pt width 4.75in
\end{table}

\subsection{LLRR vs. LLLL Operators} 
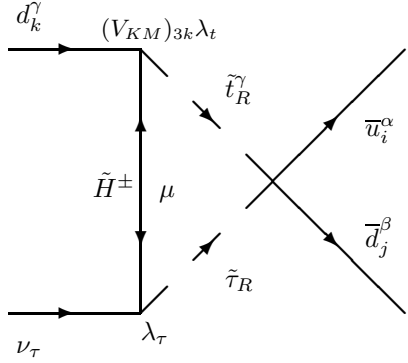
\begin{figure}
\vrule height 1pt width 4.75in
\vskip -10pt
\vrule height 1pt width 4.75in
\begin{center}
\caption{Feynman diagram that gives the dominant contribution to 
$C_{1jk3}^{(\overline {ud})(d\nu)}$.}
\label{f:2}
\begin{picture}(150,150)(0,-75)
\thicklines
\put(0,50){\line(1,0){50}}
\put(25,50){\vector(1,0){0}}
\put(0,-50){\line(1,0){50}}
\put(25,-50){\vector(1,0){0}}
\put(150,50){\line(-1,-1){50}}
\put(150,-50){\line(-1,1){50}}
\multiput(50,50)(20,-20){3}{\line(1,-1){10}}
\multiput(50,-50)(20,20){3}{\line(1,1){10}}
\put(50,50){\line(0,-1){100}}
\put(50,25){\vector(0,1){0}}
\put(50,-25){\vector(0,-1){0}}
\put(78,22){\vector(1,-1){0}}
\put(78,-22){\vector(1,1){0}}
\put(125,25){\vector(1,1){0}}
\put(125,-25){\vector(1,-1){0}}
\put(3,60){$d_k^\gamma$}
\put(3,-65){$\nu_\tau$}
\put(82,32){$\tilde t^\gamma_R$}
\put(82,-40){$\tilde\tau_R$}
\put(135,-25){$\overline d_j^\beta$}
\put(135,17){$\overline u_i^\alpha$}
\put(32,-6){$\tilde H^\pm$}
\put(57,-6){$\mu$}
\put(35,55){\small $(V_{KM})_{3k} \lambda_t$}
\put(50,-60){$\lambda_\tau$}
\end{picture}
\end{center}
\vrule height 1pt width 4.75in
\vskip -10pt
\vrule height 1pt width 4.75in
\end{figure}

Secondly, LLRR operators dominate over
LLLL operators for the third generation anti-neutrino, for the decays into
$K\overline\nu$ and $\pi\overline\nu$. (See Tables A4 and B4.) We can gain an
intuitive understanding of why this is if we neglect the gluino contribution to
the decay rates and look at approximate formulas for the rate of decay due to
charginos. The loop integral
$I(a,b,c)$ is a relatively smooth function of the masses and as a result, to a
very good approximation, when calculating chargino diagrams for the third
generation anti-neutrino, sums over gamma matrices of the form $\G_{L \;
\lam i}
\G{}^*_{R
\;
\lam j} I(\tilde\Omega_\lam,b,c)$ are approximately zero while sums of the form
$\G_{L
\;
\lam i}
\G{}^*_{L \;
\lam j} I(\tilde\Omega_\lam,b,c)$ and $\G_{R \; \lam i} \G{}^*_{R \; \lam j}
I(\tilde\Omega_\lam,b,c)$ are approximately equal to $\delta_{i j}
I(\tilde\Omega_{i_L},b,c)$ and $\delta_{i j} I(\tilde\Omega_{i_R},b,c)$,
respectively, due to the orthogonality of the gamma matrices. Hence,
\def\vkm{V_{KM}}
\begin{equation}
\label{e:5.1}
\begin{array}{rcl} C_{1jk3}^{(\overline{ud})(d\nu)} & \approx & \nn\\
\multicolumn{3}{c}{\fr 1.16 \pi^2 \mt. (\G_{U,R} \hat Y_U \vkm)_{\lam k}
\G{}^*_{U,R \; \lam k'} (\G_{E,R}
\hat Y_e)_{\rho 3} \G{}^*_{E,R \rho l'} \cudh^{[1j} \cueh^{k'] l'} U_{+
\; 2n} U_{- \; 2n} m_{\td\chi_n} I(\td\chi_n,\td u_\lam,\td e_\rho) }\nn\\
&\approx &
\fr 1.16 \pi^2 \mt.\lam_\tau \lam_{u_{k'}} (\vkm)_{k'k} \cudh^{[1j} \cueh^{k']
3} U_{+
\; 2n} U_{- \; 2n} m_{\td\chi_n} I(\td\chi_n,\td u_{k'_R},\td e_{3_R}) \nn\\ &
\approx & 
\fr 1.16 \pi^2 \mt.\lam_\tau \lam_{t} (\vkm)_{3k} \cudh^{[1j} \cueh^{3] 3} U_{+
\; 2n} U_{- \; 2n} m_{\td\chi_n} I(\td\chi_n,\td u_{3_R},\td e_{3_R})
\end{array}
\end{equation} See fig. \ref{f:2} for the Feynman diagram giving the dominant
contribution to eqn. (\ref{e:5.1}).
% Similarly,
\begin{eqnarray}
\label{e:5.2} C_{1jk3}^{(ud)({d\nu})} & \approx & \nn\\ &&\fr 1.16 \pi^2 \mt.
U_{+ \; 1n} U_{- \; 1n} m_{\td\chi_n}\{ g_2^2 (\vkm)_{i' j}
(\vkm^\dagger)_{j'1}
\cqqh^{i'[j'} \cqlh^{k]3} I(\td\chi_n,\td u_{i'_L},\td d_{j'_L})\nn
\\ && +g_2^2 (\vkm)_{k'k} \cqqh^{[1j} \cqlw^{k']3} I(\td\chi_n,\td u_{k'_L},\td
e_{3_L})\}
\end{eqnarray}

The loop integral factors $\sum_n U_{+ \; a n} U_{- \; a' n} m_{\td\chi_n}
I(\td\chi_n,b,c)$ can further be approximated
\begin{eqnarray}
\sum_n U_{+ \; a n} U_{- \; a' n} m_{\td\chi_n} I(\td\chi_n,b,c) &\approx&
\cases{ M_{wino} I(\tilde W_\pm,b,c)& if $a=a'=1$\cr
\mu_R I(\tilde H_\pm,b,c)& if $a=a'=2$}.\nn
\end{eqnarray} Hence,
\footnotesize
\begin{equation}
\hskip -.6 in
\fr C_{1jk3}^{(\overline{ud})({d\nu})}.C_{1jk3}^{(ud)({d\nu})}. \approx 
\fr \mu_R.M_{wino}. {{\lam_\tau \lam_{t} (\vkm)_{3k} \cudh^{[1j} \cueh^{3] 3}}
I(\tilde H_\pm,\td u_{3_R},\td e_{3_R})
\over {g_2^2 (\vkm)_{i' j} (\vkm^\dagger)_{j'1} \cqqh^{i'[j'} \cqlh^{k]3}
I(\tilde W_\pm,\td u_{i'_L},\td d_{j'_L})+ g_2^2 (\vkm)_{k'k} \cqqh^{[1j}
\cqlw^{k']3}}I(\tilde W_\pm,\td u_{k'_L},\td e_{3_L}) } 
\end{equation}
\normalsize The integral $I(a,b,c)$ is approximately $\log(b^2/c^2)/(b^2-c^2)$
in the limit where $a \ll b,c$, i.e. when squarks and sleptons are much more
massive than gauginos, which generally is the limit we are interested in. Using
the fact that the first and second generation squarks are approximately
degenerate, and that the first and second generation squarks  are usually more
massive than the third generation squarks and sleptons, the ratio can be
further approximated
\begin{equation}
\hskip -.6 in
\fr C_{1jk3}^{(\overline{ud})({d\nu})}.C_{1jk3}^{(ud)({d\nu})}. \approx 
\fr \mu_R.M_{wino}. \fr (m_{\rm squark}^{1,2})^2 \log(m^2_{\tilde
u_{3_R}}/m^2_{\tilde e_{3_R}}).m^2_{\tilde u_{3_R}}-m^2_{\tilde e_{3_R}}.
{{\lam_\tau \lam_{t} (\vkm)_{3k} \cudh^{[1j}
\cueh^{3] 3}}
\over {g_2^2 (\vkm)_{i' j} (\vkm^\dagger)_{j'1} \cqqh^{i'[j'} \cqlh^{k]3}
+g_2^2 (\vkm)_{k'k} \cqqh^{[1j} \cqlw^{k']3}} } 
\end{equation} where $m_{squark}^{1,2}$ is the mass of the first and second
generation squarks.

Several factors contribute to the fact that this ratio is often greater than
one.  Since the third generation squarks and sleptons are lighter than the
first and second generation squarks, the ratio is significantly enhanced by
the factor
$(m_{\rm squark}^{1,2})^2 \log(m^2_{\tilde u_{3_R}}/m^2_{\tilde
e_{3_R}})/(m^2_{\tilde u_{3_R}}-m^2_{\tilde e_{3_R}})
\approx (m_{\rm squark}^{1,2})^2/(\max(m_{\tilde u_{3_R}},m_{\tilde
e_{3_R}}))^2$. Second, the ratio 
\begin{equation}
\label{e:5.11} {{\lam_\tau \lam_{t} (\vkm)_{3k} \cudh^{[1j}
\cueh^{3] 3}}
\over {g_2^2 (\vkm)_{i' j} (\vkm^\dagger)_{j'1} \cqqh^{i'[j'} \cqlh^{k]3}
+g_2^2 (\vkm)_{k'k} \cqqh^{[1j} \cqlw^{k']3}} }
\end{equation} is itself typically of order unity for many values of the GUT
scale initial parameters.  Note, this ratio is greatly enhanced in the regime
of large
$\tan\beta$ considered in this paper.

Finally, the ratio $\mu_R/M_{wino}$ plays a critical role in whether the LLRR
operators dominate over LLLL operators in the third generation and whether the
third generation anti-neutrino dominates over the second-generation. Comparing
Tables A4, A5A, A5B, B4, B5A, and B5B with Tables A6 and B6, we see that there
is a direct correlation between
$\mu_R/m_{1/2}$ and the dominance of the third generation LLRR operators. When
$\mu_R/m_{1/2}$ is small, the third generation LLRR operators can be
suppressed. Moreover, because the third generation LLRR operators are
suppressed when
$\mu_R/m_{1/2}$ is small, the second generation anti-neutrino contributes more
significantly than the third when $\mu_R/m_{1/2}$ is small.\footnote{Arnowitt,
et al. observed that LLRR operators can be significant to nucleon decay rates
under certain circumstances in ref.
\cite{an:early}.} In particular, in runs I and II of the A and B tables,
$\mu_R/m_{1/2}$ is small ($\sim .3$) while in runs III(1), III(2), and III(3),
$\mu_R/m_{1/2}$ is near 1 or greater, and $\mu_R/m_{1/2}$ increases as one goes
from run III(1) to run III(2) to run III(3). As a result, looking at the tau
anti-neutrino columns of Tables A4 and B4, we see that, for any particular
model, the entries of those columns for runs I and II are smaller than they are
for runs III(1), III(2), and III(3), and that the entries in those columns
increase steadily in going from run III(1) to III(2) to III(3).  Similarly,
looking at the 3rd columns of Tables A5A, A5B, B5A, and B5B, we see that the
LLLL operators of the second generation anti-neutrino are fairly significant to
the overall decay rate in runs I and II, while they are not quite as
significant in runs III(1), III(2), and III(3), and that the significance of
the LLLL operators of the second generation anti-neutrino continually
decreases in going from run III(1) to III(2) to III(3).

For the first and second generation anti-neutrinos, on the other hand, the LLRR
operators are negligible because they are suppressed in comparison to the LLLL
operators by the up and charm Yukawa couplings, respectively. Thus, the entries
in the electron and muon anti-neutrino columns of Tables A4 and B4 are  fairly
small, and the entries in columns 2 and 4 of Tables A5A, A5B, B5A, and B5B are
fairly small. In comparison, the second generation anti-neutrino LLLL operator
is the most significant of the LLLL operators, but the third generation LLLL
operator is not negligible in comparison to the second generation LLLL
operator. (See columns 1, 3, and 5 of Tables A5A, A5B, B5A, and B5B.).

\subsection{$p \rightarrow \pi^+ \overline{\nu}$ vs.  $p \rightarrow K^+
\overline{\nu}$} Secondly, we see that under certain circumstances, the decay
$p
\to
\pi^+\overline\nu$ dominates over $K^+\overline\nu$ when the \co{13} operator
is included in model 4(c). When the third generation anti-neutrino dominates
over the the other generations, we have the approximate result
\def\kv{{K^+\overline\nu}}
\def\pv{{\pi^+\overline\nu}}
$$\fr\G(p\to\pv).\G(p\to\kv). \approx 3.9 \frac{\abs{(\vkm)_{31} \cudh^{[11}
\cueh^{3]3}}^2}{\abs{.28 (\vkm)_{31} \cudh^{[12} \cueh^{3]3}+(\vkm)_{32}
\cudh^{[11} \cueh^{3]3}}^2}$$
\begin{equation} = 3.9 \abs{\fr(\vkm)_{31}.(\vkm)_{32}.}^2
\frac{\abs{\cudh^{[11}
\cueh^{3]3}}^2}{\abs{.28 \fr(\vkm)_{31}.(\vkm)_{32}. \cudh^{[12}
\cueh^{3]3}+\cudh^{[11} \cueh^{3]3}}^2}
\label{e:5.x}
\end{equation}

In Appendix 4, we show that $|(\vkm)_{31}| \equiv |V_{td}|$ increases when the
\co{13} operator is included. The increase in the ratio of the rate of
$p\to\pv$ versus $p\to\kv$ when the
\co{13} operator is included can be attributed in large part to this increase
in
$|V_{td}|$, since eqn. \ref{e:5.x} contains a multiplicative factor of
$|V_{td}/V_{ts}|^2$. This increase is further enhanced by the fact that
$(\vkm)_{31} \cudh^{[12} \cueh^{3]3}$ has roughly the opposite sign of
$(\vkm)_{32} \cudh^{[11} \cueh^{3]3}$, and hence increasing
$|V_{td}|$ decreases the denominator in eqn. \ref{e:5.x}.  Thus, the addition
of the
\co{13} operator in model 4(c)  increases the ratio of the $p \to \pv$ decay
rate to $p
\to \kv$, \it provided
\rm that the third generation anti-neutrino dominates. 

Whether the third generation dominates over the second depends on the ratio of
$\mu_R/m_{1/2}$. Thus, $p \to \pv$ will be larger than $p \to \kv$ if the
\co{13} operator is included and $\mu_R/m_{1/2}$ is not much smaller than one.
Thus, in runs III(1), III(2), and III(3) of Table A1, $\mu_R/m_{1/2}$ is
approximately one or bigger, and as a result the third generation anti-neutrino
dominates the rate of decay, and the ratio of the rate of decay into $\pv$
versus the rate of decay into
$\kv$ is significantly enhanced in comparison to the runs without the \co{13}
operator in Table B1. On the other hand, in runs I and II, $\mu_R/m_{1/2}$ is
small, and as a result, the second generation dominates and the ratio of the
rate of decay into $\pv$ versus $\kv$ remains near what it was without the
\co{13} operator.

\subsection{``Generic" SU(5) vs Large $\tan\beta$ SO(10) models}
\subsubsection{$n \rightarrow \pi^0 \overline{\nu}$ vs. $n \rightarrow \eta
\overline{\nu}$} Furthermore, the tables of the previous section show some
important differences between the nucleon decay predictions for our SO(10)
model versus the predictions of a generic SUSY minimal SU(5) model. Because the
effective color triplet Higgs mass is constrained to be lower than around
$10^{17}$~GeV in SUSY minimal SU(5) \cite{hisano}, and because the lifetimes of
the nucleons are proportional to
$\sin^2 2 \beta$
\cite{an,an:early,hisano}, minimal SU(5) models use small $\tan \beta$ to be
consistent with the experimental limits on proton decay. When
$\tan
\beta$ is small, LLRR operators can often be neglected. When LLRR operators are
negligible, the ratio $\G(n \to \pi^0\overline\nu)/\G(n \to \eta
\overline\nu)$ just depends on chiral Lagrangian factors.
\begin{eqnarray}
\fr \G(n \to \pi^0\overline\nu).\G(n \to \eta \overline\nu). &\approx&
 2.8 {\sum_i\abs{\beta C^{(ud)(d\nu_i)}+\alpha
C^{(\overline{ud})(d\nu_i)}}^2\over 
\sum_i \abs{\beta C^{(ud)(d\nu_i)}-.140 \alpha C^{(\overline{ud})(d\nu_i)}}^2
}\nn\\ &\approx& 2.8
\end{eqnarray} In contrast, when $\tan\beta$ is large, LLRR operators are not
negligible. The contribution of LLRR operators to $n \to \eta \overline\nu$ is
significantly suppressed in comparison to its contribution to $n \to
\pi^0\overline\nu$ by chiral Lagrangian factors. Hence, looking at Tables A1B
and B1B, the rate of
$n \to \pi^0\overline\nu$ can be anywhere from 2.9 to over 100 times larger
than the rate of $n \to
\eta\overline\nu$, depending on whether the third generation LLRR operators or
the second generation LLLL operators dominate. 

\subsubsection{$n \rightarrow K^0 \overline{\nu}$ vs. $p \rightarrow K^+
\overline{\nu}$} Secondly, the ratio $\G(n \to K^0 \overline\nu)/\G(p \to
K^+\overline\nu)$ differs significantly from the generic SU(5) models.
Numerically,
\begin{equation}
\fr \G(n \to K^0 \overline\nu).\G(p \to K^+\overline\nu).  \approx 
\abs{ {
\beta (1.14 C^{(us)(d\nu_i)}+1.58 C^{(ud)(s\nu_i)})+\alpha (-.86
C^{(\overline{us})(d\nu_i)} +1.58 C^{(\overline{ud})(s\nu_i)})
\over
\beta (.44 C^{(us)(d\nu_i)}+1.58 C^{(ud)(s\nu_i)})+\alpha (.44
C^{(\overline{us})(d\nu_i)} +1.58 C^{(\overline{ud})(s\nu_i)}) } }^2
\end{equation} In the minimal SU(5) model, the $C^{(us)(d\nu_i)}$ operator is
approximately equal to the $C^{(ud)(s\nu_i)}$ \cite{an,an:early,hisano}, and,
since LLRR operators are often negligible, $\G(n
\to K^0 \overline\nu)/\G(p \to K^+\overline\nu) \approx 1.8$. \cite{hisano}.
However, in our SO(10) model, LLRR operators tend to dominate. Not only are the
chiral Lagrangian factors different when the LLRR operators dominate, but
$C^{(\overline{us})(d\nu_i)}$ tends to point in the \it opposite \rm direction
as $C^{(\overline{ud})(s\nu_i)}$. Hence, $\G(n
\to K^0 \overline\nu)/\G(p \to K^+\overline\nu)$ is much larger than its value
in the minimal SU(5) models. Indeed, $n \to K^0\overline\nu$ can be over 18
times bigger than $p \to K^+\overline\nu$.

Also noteworthy is the fact that when LLRR operators dominate, 
$\G(n \to K^0 \overline \nu)/\G(p \to K^+ \overline \nu)$ is significantly
higher when the \co{13} operator is included in comparison to when it is not.
(For example, in run III(3) $\G(n \to K^0 \overline \nu)/\G(p \to K^+ \overline
\nu)$ is 18.1 with the \co{13} operator included while it is no greater than
4.5 for run III(3) without the \co{13} operator.) Much of the enhancement can
be explained by the fact that $|V_{td}|$ is larger when the \co{13} operator is
included. When LLRR operators dominate,
\begin{equation}
\fr \G(n \to K^0 \overline\nu).\G(p \to K^+\overline\nu).  \approx 
\abs{ { -.86 C^{(\overline{us})(d\nu_i)} +1.58 C^{(\overline{ud})(s\nu_i)}
\over .44 C^{(\overline{us})(d\nu_i)} +1.58 C^{(\overline{ud})(s\nu_i)} } }^2
\end{equation} Plugging eqn. \ref{e:5.1} into this formula, this becomes
\begin{equation}
\fr \G(n \to K^0 \overline\nu).\G(p \to K^+\overline\nu).  \approx 
\abs{ { -.86 (\vkm)_{31} \cudh^{[12} \cueh^{3] 3} +1.58 (\vkm)_{32} \cudh^{[11}
\cueh^{3] 3}
\over .44 (\vkm)_{31} \cudh^{[12} \cueh^{3] 3} +1.58 (\vkm)_{32} \cudh^{[11}
\cueh^{3] 3} } }^2
\end{equation} Since $(\vkm)_{31} \cudh^{[12} \cueh^{3] 3}$ tends to have the
opposite sign as
$(\vkm)_{32} \cudh^{[11} \cueh^{3] 3}$,  $\G(n \to K^0 \overline \nu)/\G(p \to
K^+ \overline
\nu)$ is enhanced when $|V_{td}|$ is increased.

\subsection{Sensitivity to ``22" Clebsch} Furthermore, by looking at Tables B1
through B6, we can determine how sensitive nucleon decay rate predictions are
on the Clebsches that enter into the
$c_{qq}$, $c_{ql}$, $c_{ud}$, and $c_{ue}$ matrices. For each of the five runs
of Tables B1 through B6, the {\it only} difference between the versions $a$
through $f$ in each run are the $y_{qq}$, $y_{ql}$, $y_{ud}$ and $y_{ue}$
Clebsches of Table 1 that enter into the $c_{qq}$, $c_{ql}$, $c_{ud}$, and
$c_{ue}$ matrices. We see that the overall rate of decay is quite sensitive to
the different choices for Clebsches. For example, with
$\beta=-\alpha$, $\tau(p \to K^+ \overline\nu)$ for run I(e) is 7 times larger
than $\tau(p \to K^+ \overline\nu)$ for run I(f). The branching ratios
generally exhibit less sensitivity: $\G(n \to \pi^0 \overline\nu)/\G(n \to K^0
\overline\nu)$ exhibits virtually no sensitivity and $\G(p \to \pi^+
\overline\nu)/\G(p \to K^+ \overline\nu)$ exhibits relatively mild sensitivity.
However, branching ratios into less dominant decay modes can at certain times
exhibit high sensitivity. For example, with $\beta=-\alpha$, $\G(n \to
\eta\overline\nu)/\G(n \to K^0\overline\nu)$ is over 12 times larger for run
I(f) than it is for run I(c). 

\subsection{Gluino vs. Chargino contributions} It can also be seen that the
contributions of gluinos to the rate of nucleon decay is often not negligible.
Indeed, in several examples, excluding the gluinos' contribution can lead to a
decrease in the predicted rate of decay of greater than 60\%, in cases where
gluinos constructively interfere, or an increase in the rate of decay by over
150\%, where gluinos destructively interfere.\footnote{Goto, et al. observed
that gluino loops can be important in the minimal SU(5) model in ref.
\cite{goto}.} Note also that whether gluinos constructively or destructively
interfere depends heavily on the phase of the chiral Lagrangian parameter
$\arg(\beta/\alpha)$.

\subsection{Proton decay from gauge boson exchange} Finally we note that our
analysis only includes the contribution to nucleon decay from the effective
dimension 5 operators resulting from colored triplet Higgs exchanges. We have
neglected the contribution to  nucleon decay via heavy gauge boson exchange
(effective dimension 6 operators).  This approximation is justified in our
models for the dominant decay modes.  For example, in order to obtain the
$\ep_3$ of run III(1), we can choose the vevs
$a_1$,
$a_2$, $\tilde a$, and a singlet field
$\cs{4}$, defined in paper I, which enter into the SO(10) breaking sector of
the theory, to be $2.0 \times 10^{16}$~GeV, $1.0 \times 10^{16}$~GeV, $6.0
\times 10^{16}$~GeV, and $.66 \times 10^{16}$~GeV, respectively, i.e. all of
order
\mg.   Then the masses of the gauge bosons contained in $ SO(10)/(SU(3)\times
SU(2)\times U(1))$ -- $X^\pm$, $Q^\pm$, $U^\pm$, and
$E^\pm$ are $3.2 \times 10^{16}$~GeV, $1.2 \times 10^{17}$~GeV, $1.3 \times
10^{17}$~GeV, and $1.3 \times 10^{17}$~GeV, respectively\footnote{Note, the
$X^\pm$ gauge boson is the massive gauge boson from the 24 representation of
SU(5); $Q^\pm$ is the gauge boson from the 10, 15, $\overline {10}$, and
$\overline {15}$ representations of SU(5) which is in the $(3,2,\fr 1.3.)$ and
$(\overline 3,2,-\fr 1.3.)$ representations of SU(3)$\times$SU(2)$\times$U(1);
and $U^\pm$ and $E^\pm$ are gauge bosons from the 10 and $\overline {10}$
representations of SU(5) which are in $\{(3,1,\fr 4.3.),(\overline 3,1,-\fr
4.3.)\}$ and $\{(1,1,-2),(1,1,2)\}$ representations of
SU(3)$\times$SU(2)$\times$U(1), respectively.}. The decay mode which would be
the most dominant if all other contributions to proton decay except the
contribution due to gauge boson exchanges were neglected is $p
\to\pi^0 e^+$. With the above gauge boson masses, the partial proton lifetime
due to heavy gauge boson exchanges for $p
\to\pi^0 e^+$ is $1.2 \times 10^{38}$~yrs, corresponding to a branching ratio
of order $ < 10^{-4}$.  Of the decay modes listed in Tables A1A and B1A, gauge
exchange is competitive only with $p \to \eta\mu^+$.

\section*{Conclusions}

We have shown in this paper that model 4(c) of paper I predicts nucleon decay
rates consistent with all current experimental bounds, while using values of
GUT parameters that give fermion masses, mixing angles, and gauge couplings in
good agreement with experimental observations.  Our main results can be found
in Tables A1A, A1B, and A6.  We conclude that our model predicts that nucleon
decay is likely to be observed by SuperKAMIOKANDE or ICARUS, which are
expected to probe nucleon lifetimes up to around $10^{34}$ yrs
\cite{skam,icarus}, for various decay modes predicted by GUTs. 

  In order to avoid this conclusion one would need to make squarks and sleptons
``unnaturally" heavy and beyond the reach of LHC or
 increase the effective color triplet Higgs mass \mt, which would require a
supermassive Higgs doublet in the GUT desert with mass many orders of magnitude
lower than the GUT scale and at least an order of magnitude lighter than any
other particle getting mass around the GUT scale.   Moreover, we have chosen
the  poorly known chiral Lagrangian parameter $|\beta|$ to be at the lowest
value suggested by the data. If it could be shown that $|\beta|$ lies in the 
higher range of its current bounds, non-observation of nucleon decay by
SuperKAMIOKANDE and ICARUS could make our model unnatural for any reasonable
values of the squark and slepton masses.\footnote{ Recall, however, that the
chiral Lagrangian approach tends to overestimate nucleon decay rates
\cite{gavela} and thus underestimates the lifetimes.} For these reasons, we
believe that these models, if correct, necessarily lead to observable nucleon
decay rates.
 
We have shown that LLRR operators are not only significant, but often dominate,
nucleon decay in the large tan~$\beta$ regime -- as long as $\mu_R/m_{1/2}$ is
not very small. As a result, if nucleon decay is observed, there are two key
experimental observables that may distinguish between large or small
tan~$\beta$ SUSY GUTs: the ratios $\G(n\to K^0\overline\nu)/\G(p \to
K^+\overline\nu)$ and
$\G(n\to\eta\overline\nu)/\G(n\to\pi^0\overline\nu)$. In particular, evidence
for a neutron lifetime  $({1 \over 5} - {1 \over 20}) \times$ the proton
lifetime would be a strong indication for large tan~$\beta$ SUSY GUTs. 
Observation of $\G(n\to\eta\overline\nu)/\G(n\to\pi^0\overline\nu)$
significantly lower than than the predicted value when LLRR operators are
negligible could also indicate  large tan~$\beta$ SUSY GUTs. 

We have also shown that gaugino loops cannot be neglected when calculating
proton decay rates in models such as ours. In fact, neglecting gaugino loops
could lead to an underestimation of the decay rates of over 60\% or
overestimation of over 150\%.

\def\cmat{$c_{qq}$, $c_{ql}$, $c_{ud}$, and $c_{ue}$}
\def\ymat{$Y_u$, $Y_d$, and $Y_e$} Finally, we have studied the sensitivity of
nucleon lifetime and branching ratio predictions on the ``quality" of the
predictions that these models make for fermion masses and mixing angles. In the
models we have analyzed, the entries in the \cmat\ matrices are related to
entries in the \ymat\ matrices by Clebsches which depend on the version of the
model being considered. As we have seen, the lifetimes and branching ratios 
can be quite sensitive to the choice of Clebsches --- some predictions vary by
nearly an order of magnitude depending on the choice of Clebsches.   In models
4(a) through (f), without the \co{13} operator, the different Clebsches have no
effect on the predictions for fermion masses and mixing angles.  Comparing
models 4(a) through (f), without the \co{13} operator, which is consistent with
fermion masses and mixing angles at 2$\sigma$, with model 4(c), with the
\co{13} operator, which is consistent within 1$\sigma$, one is lead to
conclude that fitting the data within 1 or 2
$\sigma$ can have a significant effect on the nucleon decay predictions.   Thus
``predictions" for nucleon decay lifetimes and branching ratios  cannot be
expected to be any better than the complementary predictions for fermion masses
and mixing angles.

\section*{Acknowledgments} This research was supported in part by the U.S.
Department  of Energy contract DOE/ER/01545-697.  We would like to thank
Tom\'a\v s Bla\v{z}ek, Marcela Carena, and Carlos Wagner for letting us use the
results of work in progress on a general $\chi^2$ analysis of fermion masses.

\appendix

\section*{Appendix 1: How the Dimopoulos-Wilczek mechanism can produce $\mt \gg
M_{GUT}$}

Using the Dimopoulos-Wilczek mechanism for doublet-triplet splitting, the part
of the superspace potential in our model giving doublet-triplet splitting is
$$W_{d-t}=10_1 A_1 10_2 + 10_2^2 \cs{*},$$ where $A_1$ is a 45 representation
getting a vev of order \mg\ in the baryon minus lepton number direction,
$10_1$ and $10_2$ are 10 representations, and \cs{*} is a singlet. The Higgs
doublet and triplet mass matrices for our model are
\begin{equation} M_t=\bordermatrix{ & 10_1 & 10_2  \cr 10_1 & 0 & a_1 
\cr 10_2 & -a_1 & \cs{*}  \cr }\nn
\end{equation}
\begin{equation} M_d=\bordermatrix{ & 10_1 & 10_2  \cr 10_1 & 0 & 0  \cr 10_2 &
0 & \cs{*}  \cr }
\end{equation}  Note, the
$10_1$ field is the only 10 representation that couples to ordinary (Standard
Model) fermions. Thus, the $10_1$ contains the two Higgses of the Minimal SUSY
Standard Model [MSSM], and the baryon-number violating effective operators
obtained by integrating out the color triplets are proportional to
$1/\mt \equiv (M^{-1}_t)_{11}=\cs{*}/a_1^2$, the $10_1, 10_1$ entry of the
inverse of the color triplet mass matrix. An effective color triplet mass of
around
$10^{19}$~GeV is obtained with $a_1$  around the GUT scale, and \cs{*} around
$10^{13}$~GeV. Thus, $\mt \gg \mg$ means that there is an  electroweak doublet
several orders of magnitude lighter than the GUT scale, not that there is an
actual color triplet with mass greater than
$M_{Planck}$~\cite{babubarr}.

\section*{Appendix 2: Feynman diagrams}
$$\vcenter{\hbox{\begin{picture}(100,100)(-50,-50)
\thicklines
\put(-50,0){\vector(1,0){25}}
\put(-50,0){\line(1,0){50}}
\put(0,-50){\vector(0,1){25}}
\put(0,-50){\line(0,1){50}}
\multiput(0,0)(10,0){5}{\line(1,0){5}}
\put(25,0){\vector(1,0){0}}
\put(-33,10){$d^\beta_j$}
\put(18,10){$\tilde u_\lam^\alpha$}
\put(10,-31){$\tilde \chi_+^n$}
\end{picture}}} \qquad  -i \gu_{\lam j}
\delta^\alpha_\beta$$

$$\vcenter{\hbox{\begin{picture}(100,100)(-50,-50)
\thicklines
\put(-50,0){\vector(1,0){25}}
\put(-50,0){\line(1,0){50}}
\put(0,-50){\vector(0,1){25}}
\put(0,-50){\line(0,1){50}}
\multiput(0,0)(10,0){5}{\line(1,0){5}}
\put(25,0){\vector(1,0){0}}
\put(-33,10){$u^\alpha_i$}
\put(18,10){$\tilde d_\rho^\beta$}
\put(10,-31){$\tilde \chi_-^n$}
\end{picture}}} \qquad -i \gd_{\rho i}
\delta^\alpha_\beta$$

$$\vcenter{\hbox{\begin{picture}(100,100)(-50,-50)
\thicklines
\put(-50,0){\vector(1,0){25}}
\put(-50,0){\line(1,0){50}}
\put(0,-50){\vector(0,1){25}}
\put(0,-50){\line(0,1){50}}
\multiput(0,0)(10,0){5}{\line(1,0){5}}
\put(25,0){\vector(1,0){0}}
\put(-33,10){$e_j$}
\put(18,10){$\tilde \nu_i$}
\put(10,-31){$\tilde \chi_+^n$}
\end{picture}}} \qquad -i g_2 \G_{\nu \; ij} U_{+ \;1n}$$

$$\vcenter{\hbox{\begin{picture}(100,100)(-50,-50)
\thicklines
\put(-50,0){\vector(1,0){25}}
\put(-50,0){\line(1,0){50}}
\put(0,-50){\vector(0,1){25}}
\put(0,-50){\line(0,1){50}}
\multiput(0,0)(10,0){5}{\line(1,0){5}}
\put(25,0){\vector(1,0){0}}
\put(-33,10){$\nu_i$}
\put(18,10){$\tilde e_\rho$}
\put(10,-31){$\tilde \chi_-^n$}
\end{picture}}} \qquad -i \ge_{\rho i}$$

$$\vcenter{\hbox{\begin{picture}(100,100)(-50,-50)
\thicklines
\put(-25,0){\vector(-1,0){0}}
\put(-50,0){\line(1,0){50}}
\put(0,0){\vector(0,-1){25}}
\put(0,-50){\line(0,1){50}}
\multiput(0,0)(10,0){5}{\line(1,0){5}}
\put(25,0){\vector(1,0){0}}
\put(-33,10){$\overline{u}^\alpha_i$}
\put(18,10){$\tilde d_\rho^\beta$}
\put(10,-31){$\tilde \chi_+^n$}
\end{picture}}} \qquad i \gdb_{\rho i} \delta_\alpha^\beta$$

$$\vcenter{\hbox{\begin{picture}(100,100)(-50,-50)
\thicklines
\put(-25,0){\vector(-1,0){0}}
\put(-50,0){\line(1,0){50}}
\put(0,0){\vector(0,-1){25}}
\put(0,-50){\line(0,1){50}}
\multiput(0,0)(10,0){5}{\line(1,0){5}}
\put(25,0){\vector(1,0){0}}
\put(-33,10){$\overline{d}^\beta_j$}
\put(18,10){$\tilde u_\lam^\alpha$}
\put(10,-31){$\tilde \chi_-^n$}
\end{picture}}} \qquad i \gub_{\lam j} \delta_\alpha^\beta$$

$$\vcenter{\hbox{\begin{picture}(100,100)(-50,-50)
\thicklines
\put(-25,0){\vector(-1,0){0}}
\put(-50,0){\line(1,0){50}}
\put(0,0){\vector(0,-1){25}}
\put(0,-50){\line(0,1){50}}
\multiput(0,0)(10,0){5}{\line(1,0){5}}
\put(25,0){\vector(1,0){0}}
\put(-33,10){$\overline{e}_j$}
\put(18,10){$\tilde \nu_i$}
\put(10,-31){$\tilde \chi_-^n$}
\end{picture}}} \qquad i \gnub_{i j}$$

\section*{Appendix 3: Formulas for nucleon decay in terms of chiral Lagrangian
factors}
\def\frac#1#2{{#1\over #2}} Using the chiral Lagrangian techniques of ref.
\cite{chadha}, the rates of nucleon decay are the following.
\begin{eqnarray}
\Gamma(p \to K^+\bar{\nu}_i)
            &=& \frac{(m_p^2-m_K^2)^2}{32\pi m_p^3 f_{\pi}^2} A_L^2 
\left|[\beta C^{(us)(d\nu_i)}+\alpha C^{(\overline{us})(d\nu_i)}]
\frac{2m_p}{3m_B}D \right.\nn\\ &&\left. +[\beta C^{(ud)(s\nu_i)}+\alpha
C^{(\overline{ud})(s\nu_i)}] [1+\frac{m_p}{3m_B}(D+3F)]\right|^2\nn\\
\Gamma(p \to \pi^+\bar{\nu}_i)
           &=& \frac{m_p}{32\pi f_{\pi}^2}  A_L^2 
                \left|[\beta C^{(ud)(d\nu_i)}+\alpha
C^{(\overline{ud})(d\nu_i)}] (1+D+F)\right|^2\nn\\
\Gamma (p \to \eta e_i^+) &=&{3(m_p^2-m_\eta^2)^2\over 64 \pi f_\pi^2 m_p^3}
A_L^2 \bigl
\{ \abs{\beta C^{(ud)(ue_i)} [1+{1\over3} (3 F-D)]  -\alpha
C^{(\overline{ud})(ue_i)} {1\over 3} [1-(3 F-D)]}^2 \bigr.\nn\\ &&\bigl.
+\abs{\beta C^{(\overline{ud})(\overline{\vphantom{d}ue_i})} [1+{1\over3} (3
F-D)]  -\alpha C^{(ud)(\overline{ue_i})} {1\over 3} [1-(3 F-D)]}^2 \bigr\}\nn\\
\Gamma (p \to K^0 e_i^+) &=& {(m_p^2-m_K^2)^2\over 32 \pi f_\pi^2 m_p^3}  A_L^2
\bigl\{
\abs{\beta C^{(us)(ue_i)} [1-{m_p\over m_B} (D-F)] -\alpha
C^{(\overline{us})(ue_i)}[1+{m_p\over m_B} (D-F)]}^2 \bigr.\nn\\ &&\bigl.
+\abs{\beta C^{(\overline{us})(\overline{ue_i})} [1-{m_p\over m_B} (D-F)]
-\alpha C^{(us)(\overline{ue_i})}[1+{m_p\over m_B} (D-F)}^2 \bigr\}\nn\\
\Gamma(p \rightarrow \pi^0 e_i^+)
           &=& \frac{m_p}{64\pi f_{\pi}^2} A_L^2 \bigl\{
\abs{[\beta C^{(ud)(ue_i)}+\alpha C^{(\overline{ud})(ue_i)}]
(1+D+F)}^2\bigr.\nn\\ &&\bigl.+\abs{[\beta
C^{(\overline{ud})(\overline{\vphantom{d}ue_i})}+\alpha
C^{(ud)(\overline{ue_i})}] (1+D+F)}^2\bigr\}\nn\\
\G(n \to K^0 \overline\nu_i)&=&
\fr (m_n^2-m_K^2)^2. 32 \pi m_n^3 f_\pi^2. A_L^2\left|
\beta C^{(us)(d\nu_i)} (1-\fr m_n.3 m_B. (D-3 F)) - \alpha
C^{(\overline{us})(d\nu_i)} (1+\fr m_n.3 m_B. (D-3 F))\right.\nn\\ &&\left.
+(\beta C^{(ud)(s\nu_i)}+\alpha C^{(\overline{ud})(s\nu_i)}) (1+\fr m_n.3 m_B.
(D+3 F))\right|^2\nn\\
\G(n \to \pi^0 \overline\nu_i)&=&\fr m_n^3.64 \pi f_\pi^2. A_L^2
\abs{\beta C^{(ud)(d\nu_i)}+\alpha C^{(\overline{ud})(d\nu_i)} }^2
(1+D+F)^2\nonumber\\
\G(n \to \eta \overline\nu_i)&=& \fr 3 (m_n^2-m_\eta^2)^2. 64 \pi m_n^3
f_\pi^2. A_L^2 \abs{\beta C^{(ud)(d\nu_i)} (1+\fr 1.3. (3 F-D))-\alpha
C^{(\overline{ud})(d\nu_i)} \fr 1.3. (1+D-3 F)}^2\nn\\
\G(n \to \pi^- e^+_i)&=&\fr m_n. 32 \pi f_\pi^2. A_L^2\{\abs{\beta
C^{(ud)(ue_i)}+\alpha C^{(\overline{ud})(ue_i)}}^2+\abs{\beta
C^{(\overline{ud})(\overline{\vphantom{d}ue_i})}+\alpha
C^{(ud)(\overline{ue_i})}}^2\}(1+D+F)^2\nn
\end{eqnarray} where $m_B$ is an average Baryon mass satisfying $m_B \approx
m_\Sigma \approx m_\Lambda$ and all other notation follows
\cite{chadha}\footnote
{$\cjkl^{(ud)(d\nu)}=\cjkl^{(ud)(d\nu)[G]}+\cjkl^{(ud)(d\nu)[W]}$, etc.}. Here,
all coefficients of four-fermion operators are evaluated at
$M_Z$.
$A_L$ takes into account renormalization from $M_Z$ to 1 GeV, and is
approximately equal to .22. \cite{al}. These formulas reduce to the chiral
Lagrangian formulas given in ref. \cite{hisano} for $\beta \ne 0$ when
$\alpha=0$. In the calculations, we take
$D=.81$,
$F=.44$ \cite{hisano}, and $f_\pi=139$~MeV \cite{chadha}.

\section*{Appendix 4: Why $|V_{td}|$ increases when the \co{13} is included in
model 4(c)}

It can be seen that our $Y_d$ and $Y_e$ matrices, which have the  general form,
$$\pmatrix{ 0 & z C & u D e^{i \delta} \cr z C & y E e^{i \phi} & x B \cr u' D
e^{i \delta} & x' B & A}$$ with $C, D \ll B, E \ll A$, can be diagonalized by
multiplying the matrices on the left and right by matrices
$S$ and
$T$, respectively, where
$$S \approx \pmatrix{ 1 &-(z C-{x' u B D\over A} e^{i \delta}) \fr e^{-i
\phi}.y E. &
 -\fr u D e^{i \delta}.A.+
\fr x B e^{-i \phi}.y E A. (z C-{x' u B D\over A} e^{i \delta}) \cr  (z C-{x' u
B D\over A} e^{-i \delta}) \fr e^{i \phi}.y E. & 1 & -\fr x B.A.
\cr
\fr u D e^{-i \delta}.A.& \fr x B.A. & 1}$$
$$T\approx\pmatrix{ 1& (z C-\fr x u' B D.A. e^{-i \delta}) \fr e^{i \phi}.y E.
& 
\fr u' D e^{-i \delta}.A.\cr  -(z C-\fr x u' B D.A. e^{i \delta}) \fr e^{-i
\phi}.y E. & 1 & \fr x' B.A.
\cr -\fr u' D e^{i \delta}.A.+
\fr x' B e^{-i \phi}.y E A. (z C-\fr x u' B D.A. e^{i \delta}) & -\fr x' B.A.
&1}$$ Our $Y_u$ matrix, which has the general form
$$\pmatrix{ 0 & z C & u D e^{i \delta} \cr z C & 0 & x B \cr u' D e^{i \delta}
& x' B & A}$$ with $C, D \ll B, E \ll A$ and $C \ll x x' B^2/A$, is
diagonalized by the $S$ and
$T$ matrices
$$S\approx\pmatrix{ 1 & \FR A. x x' B^2. (z C-\fr x' u B D.A. e^{i \delta}) &
-\FR z C.x' B. \cr -\FR A. x x' B^2. (z C-\fr x' u B D.A. e^{-i \delta}) & 1 & 
-\fr x B.A. \cr
\fr u D.A. e^{-i \delta} & \fr x B.A. & 1}$$
$$T\approx\pmatrix{ 1 & -\FR A. x x' B^2. (z C-\fr x u' B D.A. e^{-i \delta}) 
&
\fr u' D.A. e^{-i \delta} \cr
\FR A. x x' B^2. (z C-\fr x u' B D.A. e^{i \delta}) & 1 &  \fr x' B.A. \cr -\FR
z C.x B. & -\fr x' B.A. & 1}$$ Therefore, at \mg
$$V_{td} \approx$$
$$\fr D.A. e^{i \delta} \, (u_u-u_d)+
\fr B C.A E. e^{-i \phi}\, \fr z_d.y_d.\, (x_d-x_u) +(\fr B.A.)^2\, \fr D.E.\,
\fr (x_u-x_d) x_d' u_d. y_d.\, e^{i(\delta-\phi)}$$
\begin{eqnarray} & = & -2\, \fr D.A.\, e^{i \delta}-12\, \fr B C.A E.\, e^{-i
\phi}+\fr 16.243.\, (\fr B.A.)^2\, \fr D.E.\, e^{i(\delta-\phi)}\nn\\ & \approx
& -12 e^{-i \phi} \ (\fr B C.A E.+\fr 1.6.\, \fr D.A.\, e^{i(\delta+\phi)})
\end{eqnarray} Since $\delta+\phi \approx 30^\circ$, the $D
e^{i(\delta+\phi)}/(6 A)$ term will increase $|V_{td}|$ provided that the $B
C/(AE)$ term does not decrease too much when the \co{13}\ operator is included.

What effect does the \co{13}\ operator have on the $A$, $B$, $C$, and $E$
parameters? The $A$ parameter will not be affected at all because $A$ is fixed
purely by the third generation masses $m_b$ and $m_\tau$ \cite{adhrs}. 
Moreover, the first generation Yukawa coupling $\lam_e$ of the $Y_e$ matrix is
approximately equal to 
\begin{equation}
\fr 243.E. \abs{C^2-\fr 109.27. C D \fr B.A. e^{i \delta}}
\end{equation} at \mg. Since $\delta$ is typically close to $2 \pi$, the term
of the absolute value linear in $D$ decreases the eigenvalue. Therefore, $C$
must be increased to compensate for the \co{13}\ operator.

On the other hand, when $B$ is evaluated by using the global $\chi^2$ analysis
of ref. \cite{blazek}, $B$ actually decreases when the \co{13}\ operator is
included. It will be shown in ref. \cite{blazek} that when a global $\chi^2$
analysis is done, $B_K$, the bag constant which comes into the theoretical
formula for the experimental observable $\epsilon_K$ measuring CP violation,
and
$|V_{cb}|$ come out too high; and
$|V_{ub}/V_{cb}|$ comes out too low, in comparison with their experimental
measurements, for model 4 without an \co{13}\ operator. Since $V_{cb} \approx
\zeta |x_d-x_u| B/A$ where
$\zeta$ is a renormalization group factor, this means that $B$ is too high. It
will also be shown in ref. \cite{blazek} that unless an \co{13}\ operator is
included, $V_{cb}$ and $V_{ub}/V_{cb}$ cannot be corrected by changing the
parameters without further increasing $B_K$, which is already too high.
Furthermore, $B$ and $E$ are related by the equation
$$\abs{3 A E e^{i \phi}-B^2} \approx \lam_\mu \lam_\tau.$$ Since
$\hbox{Re}\,\phi<0$, $E$ is lower when the \co{13}\ operator is not included
because
$B$ is higher than it should be.

However, when the \co{13}\ operator of paper I is added to model 4(c), it is
possible to lower $B$ to bring it in line with the experimental value for
$|V_{cb}|$ while simultaneously having reasonable values for $B_K$ as well as
for the other observables \cite{blazek}. The net effect of including the
\co{13}\ operator to model 4(c) is that $B$ is typically decreased by $\sim
10\%$, $E$ is increased by $\sim 18\%$, and $C$ is increased by $\sim 25\%$.
Therefore, $B C/(A E)$ actually decreases slightly
$(\sim 3\%)$. However, the decrease in that term is more than compensated for
by the increase in $|V_{td}|$ due to the $\fr D.6 A. e^{i (\delta+\phi)}$
term. The net result of the inclusion of the \co{13}\ operator is an increase
in
$|V_{td}|$ typically of $\sim 11\%$.

\end{document}